\documentclass[useAMS,usenatbib]{mn2e}
\usepackage{amsmath}
\usepackage{amssymb}
\usepackage{graphicx}
\usepackage{pdflscape}
\usepackage{threeparttable}
\usepackage{xcolor} 

\title[Machine Learned Planet Validation]{Exoplanet Validation with Machine Learning: 50 new validated Kepler planets}

\author[Armstrong et. al.]{
\parbox{\textwidth}{David J. Armstrong,$^{1,2}$\thanks{d.j.armstrong@warwick.ac.uk}
Jevgenij Gamper,$^{3,4}$
Theodoros Damoulas$^{4,5,6}$
}
\vspace{0.4cm}\\
\parbox{\textwidth}{
$^{1}$Department of Physics, University of Warwick, Gibbet Hill Road, Coventry, CV4 7AL, UK\\
$^{2}$Centre for Exoplanets and Habitability, University of Warwick, Gibbet Hill Road, Coventry CV4 7AL, UK\\
$^{3}$Mathematics of Systems CDT, University of Warwick, Gibbet Hill Road, Coventry CV4 7AL, UK\\
$^{4}$Department of Computer Science, University of Warwick, Gibbet Hill Road, Coventry CV4 7AL, UK\\
$^{5}$Department of Statistics, University of Warwick, Gibbet Hill Road, Coventry CV4 7AL, UK\\
$^{6}$The Alan Turing Institute, London, UK\\
}}

\newcommand{\mytilde}{\raise.17ex\hbox{$\scriptstyle\mathtt{\sim}$}}
\newcommand{\Kepler}{\emph{Kepler} }
\newcommand{\TESS}{\emph{TESS} }

\begin{document}
\date{Accepted . Received}

\pagerange{\pageref{firstpage}--\pageref{lastpage}} \pubyear{2002}

\maketitle

\begin{abstract}
Over 30\% of the \mytilde 4000 known exoplanets to date have been discovered using `validation', where the statistical likelihood of a transit arising from a false positive (FP), non-planetary scenario is calculated. For the large majority of these validated planets calculations were performed using the \texttt{vespa} algorithm \citep{Morton:2016ka}. Regardless of the strengths and weaknesses of \texttt{vespa}, it is highly desirable for the catalogue of known planets not to be dependent on a single method. We demonstrate the use of machine learning algorithms, specifically a gaussian process classifier (GPC) reinforced by other models, to perform probabilistic planet validation incorporating prior probabilities for possible FP scenarios. The GPC can attain a mean log-loss per sample of 0.54 when separating confirmed planets from FPs in the \Kepler threshold crossing event (TCE) catalogue. Our models can validate thousands of unseen candidates in seconds once applicable vetting metrics are calculated, and can be adapted to work with the active \TESS mission, where the large number of observed targets necessitates the use of automated algorithms. We discuss the limitations and caveats of this methodology, and after accounting for possible failure modes newly validate 50 \Kepler candidates as planets, sanity checking the validations by confirming them with \texttt{vespa} using up to date stellar information. Concerning discrepancies with \texttt{vespa} arise for many other candidates, which typically resolve in favour of our models. Given such issues, we caution against using single-method planet validation with either method until the discrepancies are fully understood.
\end{abstract}


\section{Introduction}

Our understanding of exoplanets, their diversity and population has been in large part driven by transiting planet surveys. Ground based surveys \citep[e.g.][]{Bakos:2002kc,Pollacco:2006gb,Pepper:2007ja,Wheatley:2017dm} set the scene and discovered many of the first exoplanets. Planet populations, architecture and occurrence rates were exposed by the groundbreaking \Kepler mission \citep{Borucki:2016hn}, which to date has discovered over 2300 confirmed or validated planets, and was succeeded by its follow-on K2 \citep{Howell:2014ju}. Now the \TESS mission \citep{Ricker:2015ie} is surveying most of the sky, and is expected to at least double the number of known exoplanets.

The planet discovery process has a number of distinct steps, which have evolved with the available data. Surveys typically produce more candidates than true planets, in some cases by a large factor. FP scenarios produce signals that can mimic that of a true transiting planet \citep{Santerne:2013hd,Cabrera:2017fo}. Key FP scenarios include various configurations of eclipsing binaries, both on the target star and on unresolved background stars, which when blended with light from other stars can produce eclipses very similar to a planet transit. Systematic variations from the instrument, cosmic rays or temperature fluctuations can produce apparently significant periodicities which are potentially mistaken for small planets, especially at longer orbital periods \citep{Burke:2019js,Thompson:2018gm,Burke:2015kf}.

Given the problem of separating true planetary signals from FPs, vetting methods have been developed to select the best candidates to target with often limited follow-up resources \citep{Kostov:2019ij} . Such vetting methods look for common signs of FPs, including secondary eclipses, centroid offsets indicating background contamination, differences between odd and even transits and diagnostic information relating to the instrument \citep{Twicken:2018ca}. Ideally, vetted planetary candidates are observed with other methods to confirm an exoplanet, often detecting radial velocity variations at the same orbital period as the candidate transit \citep[e.g.][]{Cloutier:2019cn}.

With the advent of the \Kepler data, a large number of generally high quality candidates became available, but in the main orbiting faint host stars, with V magnitude $>14$. Such faint stars preclude the use of radial velocities to follow-up most candidates, especially for long period low signal-to-noise cases. At this time vetting methodologies were expanded to attempt planet `validation', the statistical confirmation of a planet without necessarily obtaining extra data \citep[e.g.][]{Morton:2011eu}. Statistical confirmation is not ideal compared to using independent discovery techniques, but allowed the `validation' of over 1200 planets, over half of the \Kepler discoveries, either through consideration of the `multiplicity boost' or explicit consideration of the probability for each FP scenario. Once developed, such methods proved useful both for validating planets and for prioritising follow-up resources, and are still in use even for bright stars where follow-up is possible \citep{Quinn:2019cr,Vanderburg:2019di}.

There are several planet validation techniques in the literature: \texttt{PASTIS} \citep{Santerne:2015bb,Diaz:2014kd}, \texttt{BLENDER} \citep{Torres:2015ix}, \texttt{vespa} \citep{Morton:2011eu,Morton:2012bv,Morton:2016ka}, the newly released TRICERATOPS \citep{2020arXiv200200691G} and a specific consideration of \Kepler's multiple planetary systems \citep{Lissauer:2014ki,Rowe:2014jq}. Each has strengths and weaknesses, but only \texttt{vespa} has been applied to a large number of candidates. This dependence on one method for \mytilde 30\% of the known exoplanets to date introduces risks for all dependent exoplanet research fields, including in planet formation, evolution, population synthesis and occurrence rates. In this work we aim to introduce an independent validation method using machine learning techniques, particularly a gaussian process classifier (GPC).

Our motivation for creating another validation technique is threefold. First, given the importance of designating a candidate planet as 'true' or 'validated', independent methods are desirable to reduce the risk of algorithm dependent flaws having an unexpected impact. Second, we develop a machine learning methodology which allows near instant probabilistic validation of new candidates, once lightcurves and applicable metadata are available. As such our method could be used for closer to real time target selection and prioritisation. Lastly, much work has been performed recently giving an improved view of the \Kepler satellite target stars through GAIA, and in developing an improved understanding of the statistical performance and issues relating to \Kepler discoveries \citep[e.g.][]{Bryson:tz,Burke:2019js,Mathur:2017fh,Berger:2018il}. We aim to incorporate this new knowledge into our algorithm and so potentially improve the reliability of our results over previous work, in particular in the incorporation of systematic non-astrophysical FPs.

We initially focus on the \Kepler dataset with the goal of expanding to create a general code applicable to \TESS data in future work. Due to the speed of our method we are able to take the entire threshold crossing event (TCE) catalogue of Kepler candidates \citep{Twicken:2016ea} as our input, as opposed to the typically studied \Kepler objects of interest (KOIs) \citep{Thompson:2018gm}, in essence potentially replacing a large part of the planet detection process from candidate detection to planet validation. 

Past efforts to classify candidates in transit surveys with machine learning have been made, using primarily random forests \citep{McCauliff:2015fb,Armstrong:2018ey,Schanche:2018kl,Caceres:2019ju} and convolutional neural nets \citep{Shallue:2018jy,Ansdell:2018dq,Dattilo:2019ht,Chaushev:2019gx,Yu:2019ba,Osborn:2019wo}. To date these have all focused on identifying FPs or ranking candidates within a survey. We build on past work by focusing on separating true planets from FPs, rather than just planetary candidates, and in doing so probabilistically to allow planet validation.

Section \ref{sectFramework} describes the mathematical framework we employ for planet validation, and the specific machine learning models used. Section \ref{sectData} defines the input data we use, how it is represented, and how we define the training set of data used to train our models. Section \ref{sectModels} describes our model selection and optimisation process. Section \ref{sectVal} describes how the outputs of those models are converted into posterior probabilities, and combined with a priori probabilities for each FP scenario to produce a robust determination of the probability that a given candidate is a real planet. Section \ref{sectResults} shows the results of applying our methodology to the Kepler dataset, and Section \ref{sectDiscuss} discusses the applicability and limitations of our method, as well as its potential for other datasets.

\section{Framework}
\label{sectFramework}

\subsection{Overview}
Consider training dataset $\mathcal{D} = \{\mathbf{x}_n,s_n\}_{n=1}^N$ containing $N$ TCEs and $\mathbf{x}_n \in \mathbb{R}^d$ the feature vector of vetting metrics and parameters derived from the \textit{Kepler} pipeline. Let  $p(\mathbf{X}, \mathbf{s})$ be the joint density, of the feature array $\mathbf{X}$, and the generative hypothesis labels $s$ where $\mathbf{s}$ is the array of labels (i.e. planet, or FPs such as an eclipsing binary or hierarchical eclipsing binary). Generative modelling of the joint density has been the approach taken in the previous literature for exoplanet validation, see for example \texttt{PASTIS} \citep{Diaz:2014kd,Santerne:2015bb} where the generative probability for hypothesis label $s$ has been explicitly calculated using Bayes formula.




The scenarios in question represent the full set of potential astrophysical and non-astrophysical causes of the observed candidate signal. Let $P(s | \text{I})$ represent the empirical prior probability that a given scenario $s$ has to occur, where $s=1$ represents a confirmed planet and $s=0$ refers to the FP hypothesis, including all astrophysical and non-astrophysical FP situations which could generate the observed signal. $I$ refers to \textit{a priori} available information on the various scenarios. 

We implement several machine learning classification models  $\mathcal{M}$ discussed in Section \ref{sectModels}, with their respective parameters $\mathbf{w}_{\mathcal{M}}$. The approaches we take typically estimate the posterior predictive probability $p(s=1 | \mathbf{x}^*, \mathcal{D}, \mathcal{M})$ for an unseen feature vector $\mathbf{x}^{*}$ directly as the result of the classification algorithm. We then obtain the scenario posterior probability $p(s =1 | \mathbf{x}^*, \text{I})$ by re-weighting using the estimated empirical priors:

\begin{equation}
\label{eqnmain2}
p(s=1 | \mathbf{x}^*, \text{I}) = \frac{p(s = 1 | \mathbf{x}^*, \mathcal{D}, \mathcal{M})P(s = 1 | \text{I})}{\sum_{s} p(s | \mathbf{x}^*, \mathcal{D}, \mathcal{M})P(s | \text{I})}
\end{equation}

where the posterior predictive probability of interest $p(s = 1 | \mathbf{x}^*, \mathcal{D}, \mathcal{M})$ is given by:

\begin{equation}
\label{eqnmain3}
\int p(s = 1 | \mathbf{x}^*, \mathbf{w}_{\mathcal{M}}, \mathcal{M}) p(\mathbf{w}_{\mathcal{M}}|\mathcal{D},\mathcal{M})d\mathbf{w}_{\mathcal{M}}
\end{equation}

and $p(\mathbf{w}_{\mathcal{M}}|\mathcal{D},\mathcal{M})$ is the parameter posterior for parametric models that is typically approximated in Bayesian classification models with an approximating family. Going forwards $\mathcal{D}, \mathcal{M}$ will be dropped from our notation for clarity. 

For non-Bayesian parametric methods the marginal is completely replaced by a point estimate $\hat{\mathbf{w}}_{\mathcal{M}}$ resulting to $p(s = 1 | \mathbf{x}^*, \hat{\mathbf{w}}_{\mathcal{M}})$ and the scenario conditional as:

\begin{equation}
    \label{eqnmain4}
p(s=1 | \mathbf{x}^*, \text{I}) = \frac{p(s = 1 | \mathbf{x}^*, \hat{\mathbf{w}}_{\mathcal{M}})P(s = 1 | \text{I})}{\sum_{s} p(s | \mathbf{x}^*, \hat{\mathbf{w}}_{\mathcal{M}})P(s | \text{I})}
\end{equation}

The prior information $I$ represents the overall probability for a given scenario to occur in the \Kepler dataset, as well as the occurrence rates of planets or binaries as a whole given the \Kepler precision and target stars. In this work $I$ will also include centroid information determining the chance of background resolved or unresolved sources being the source of a signal. This approach allows us to easily vary the prior information given centroid information specific to a target which isn't otherwise available to the models. We discuss the $P(s | \text{I})$ priors in detail in Section \ref{sectPriors}.
 
 Prior factors dependent on an individual candidate's parameters, including for example the specific occurrence rate of planets at the implied planet radius, as opposed to that on average for the whole \textit{Kepler} sample, as well as the difference in probability of eclipse for planets or stars at a given orbital period and stellar or planetary radius, are incorporated directly in the model output $ p(s = 1 | \mathbf{x}^*)$.
 
\subsection{Gaussian Process Classifier}
\label{sectGPC}
A commonly used set of machine learning tools are defined through parametric models such that a function describing the process belongs to a specific family of functions i.e. linear or quadratic linear regression with a finite number of parameters. A more flexible alternative are Bayesian non-parametric models \citep{williams2006gaussian}, and specifically Gaussian Processes, where one places a prior distribution over functions $\mathbf{f}$ rather than a distribution over parameters $\mathbf{w}$ of a function. We can specify a mean function value given the inputs and a kernel function that specifies the covariance of the function between two input instances. 




In the classification setting the posterior over these latent functions $p(\mathbf{f})$ is not available in closed form and approximations are needed. The probability of interest can be computed from the approximate posterior with Monte Carlo estimates of the following integral:


\begin{equation}
    P(s=1 | \mathbf{x}^{*}) = \int \int p(s=1 | f^{*}) p(f^{*}| \mathbf{f}, \mathbf{x}^{*}) p(\mathbf{f}) d\mathbf{f} df^{*}
\end{equation}


where $f^*$ is the evaluation of the latent function $f$ on the test data point $\mathbf{x}^{*}$ for which we are predicting the label $s$. Note we have dropped $\mathcal{D}$ and $\mathcal{M}$. The first term in the integrand is the predictive likelihood function, the second term is the latent predictive density, and the final term is the posterior density over the latent functions. In classification we resort to specific deterministic approximations based on stochastic variational inference that are implemented in the \texttt{gpflow python} package \citep{deGMatthews:2017tm}. We also utilise an `inducing points' methodology whereby the large dataset is represented by a smaller number of representative points, which speeds computation and guards against overfitting. The number of such points is one of the optimised parameters. For an extensive introduction to GPs refer to \cite{williams2006gaussian, blei2017variational}.


\subsection{Random Forest \& Extra Trees}
Random Forests \citep[RFs,][]{breiman2001random} are a well-known machine learning method with several desirable properties, and history in performing exoplanet transit candidate vetting \citep{McCauliff:2015fb}. They are robust to uninformative features, allow control of overfitting, and allow measurement of the feature importances driving classification decisions. RFs are constructed using a large number of decision trees, each of which gives a classification decision based on a random subset of the input data. To keep this work as concise as possible we direct the interested reader to detailed descriptions elsewhere \citep{breiman2001random, louppe2014understanding}.

Extra Trees (ET) also known as Extremely Randomized Trees are intuitively similar in construction to Random Forest \citet{geurts2006extremely}. The only fundamental difference from RF is the feature split, where RFs perform feature splitting based on a deterministic measure such as the Gini Impurity, the feature split in an ET is random. 


\subsection{Multilayer Perceptron}

A standard linear regression or classification model is based on a linear combination of instance features passed through an activation function, with non-linearity in case of classification or identity in case of regression. A multilayer perceptron on the other hand is a set of linear transformations followed by an activation function, where the number of transformations implies the number of hidden units. Each linear transformation consists of a set number of linear combinations commonly referred to as neurons, where every neuron takes as input a linear combination from every other neuron in the previous hidden unit. The number of hidden units, neurons and activation function are hyper-parameters to choose. The interested reader should refer to \citet{bishop2006pattern} for a more in depth discussion of neural networks. 

\section{Input Data}
\label{sectData}
We use Data Release 25 (DR25) of the \Kepler data, covering quarters 1 to 17 \citep{Twicken:2016ea,Thompson:2018gm}. The data measures stellar brightness for near 200000 stars for a period of four years. Data and metadata were obtained from the NASA Exoplanet Archive \citep{Akeson:2013hr}. The \Kepler data is passed through the \Kepler data processing pipeline \citep{Jenkins:2010dh,Jenkins:2017vv}, and detrended using the Presearch Data Conditioning pipeline \citep{Stumpe:2012bj,Smith:2012ji}. Planetary candidates are identified by the Transiting planet search part of the \Kepler pipeline, which produces TCEs where candidate transits appear with a significance $>7.1\sigma$. The recovery rate of planets from this process is investigated in detail in \citet{Christiansen:vw} and \citet{Burke:tk}. These TCEs were then designated as Kepler Objects of Interest (KOIs) if they passed several vetting checks known as the `Data Validation' (DV) process detailed in \citet{Twicken:2018ca}. KOIs are further labelled as FPs or planets based on a combination of methods, typically either individual follow-up with other planet detection methods, the detection of transit timing variations \citep[e.g.][]{Panichi:2019im} or statistical validation via a number of published methods \citep[e.g.][]{Morton:2016ka}.

\subsection{Metadata}
We utilise the TCE table for \Kepler DR25 \citep{Twicken:2016ea}. This table contains 34032 TCEs, with information on each TCE as well as the results of several diagnostic checks. `Rogue' TCEs which were the result of a previous bug in the transit search and flagged using the `tce\_rogue\_flag' column were removed, leaving 32534 TCEs for this study which form the basis of our dataset.

We update the TCE table with improved estimates of stellar temperature, surface gravity, metallicity and radius derived using Gaia DR2 information \citep{Berger:2018il,Collaboration:2018io}. In each case, if no information is available for a given \Kepler target in \citet{Berger:2018il}, we fall back on the values in \citet{Mathur:2017fh}, and in cases with no information in either use the original values in the TCE table, which are from the Kepler Input Catalogue \citep[KIC, ][]{Brown:2011dr}. We also include \Kepler magnitudes from the KIC. The planetary radii are updated in line with the updated stellar radii. We also recalculate the maximum ephemeris correlation, a measure of correlation between TCEs on the same stellar target\citep{McCauliff:2015fb} and add it to the TCE table.

One element of the TCE table is several $\chi^2$ and degrees of freedom statistics for various models fitted to the TCE signal. To better represent this test, we convert all such columns into the ratio of the $\chi^2$ to the degrees of freedom. Missing values are filled with their column median in the case of stellar magnitudes, or zeros for all other columns.

The full range of included data is shown in Table \ref{tabfeatures}. This is a subset of the original TCE table, with several columns removed based on their contribution to the models as described in Section \ref{sectfeatimp}. Brief descriptions of each column are given, readers should refer to the NASA Exoplanet Archive for further detail.

\subsection{Lightcurves}
\begin{figure}
\resizebox{\hsize}{!}{\includegraphics{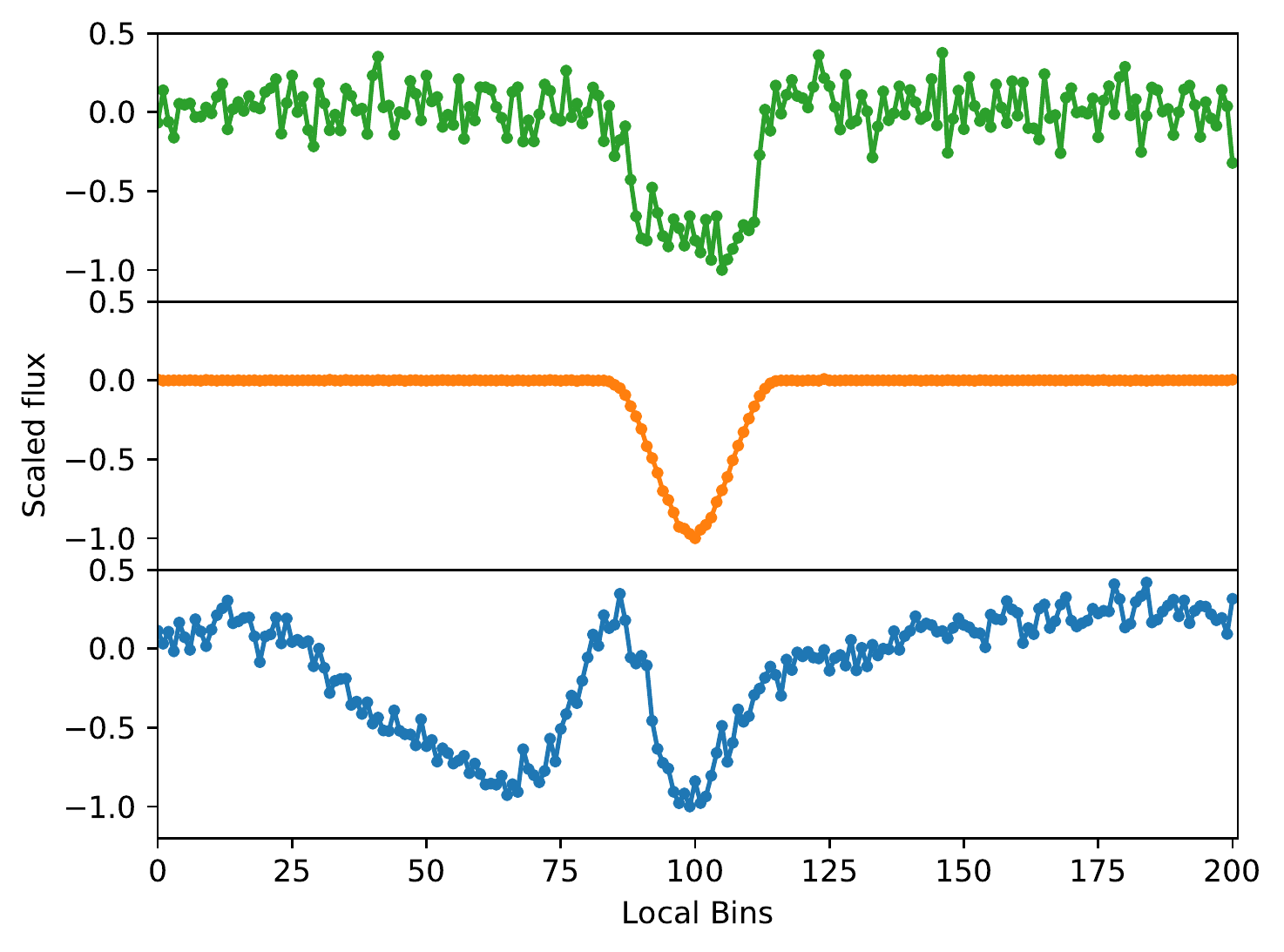}}
\caption{`Local view' 201 bin representation of the transit for a planet (top), astrophysical FP (middle) and non-astrophysical FP (bottom).}
\label{figlocview}
\end{figure}

We use the DV \Kepler lightcurves, as detailed in \citet{Twicken:2018ca}, which are produced in the same way as lightcurves used for the \Kepler Transiting Planet Search (TPS). The lightcurve data is phasefolded at the TCE ephemeris then binned into 201 equal width bins in phase covering a region of seven transit durations centred on the candidate transit. We choose these parameters following \citet{Shallue:2018jy}, their `local' view, although we use a window covering one less transit duration to provide better resolution of the transit event. Example local views are shown in Figure \ref{figlocview}. Empty bins are filled by interpolating surrounding bins. As in \citet{Shallue:2018jy} we also implemented a `global' view using 2001 phase bins covering the entire phase-folded lightcurve, but in our case found no improvement in classifier performance and so dropped this view to reduce the input feature space. We hypothesise that this is due to the inclusion of additional metrics measuring the significance of secondary eclipses.

\begin{figure}
\resizebox{\hsize}{!}{\includegraphics{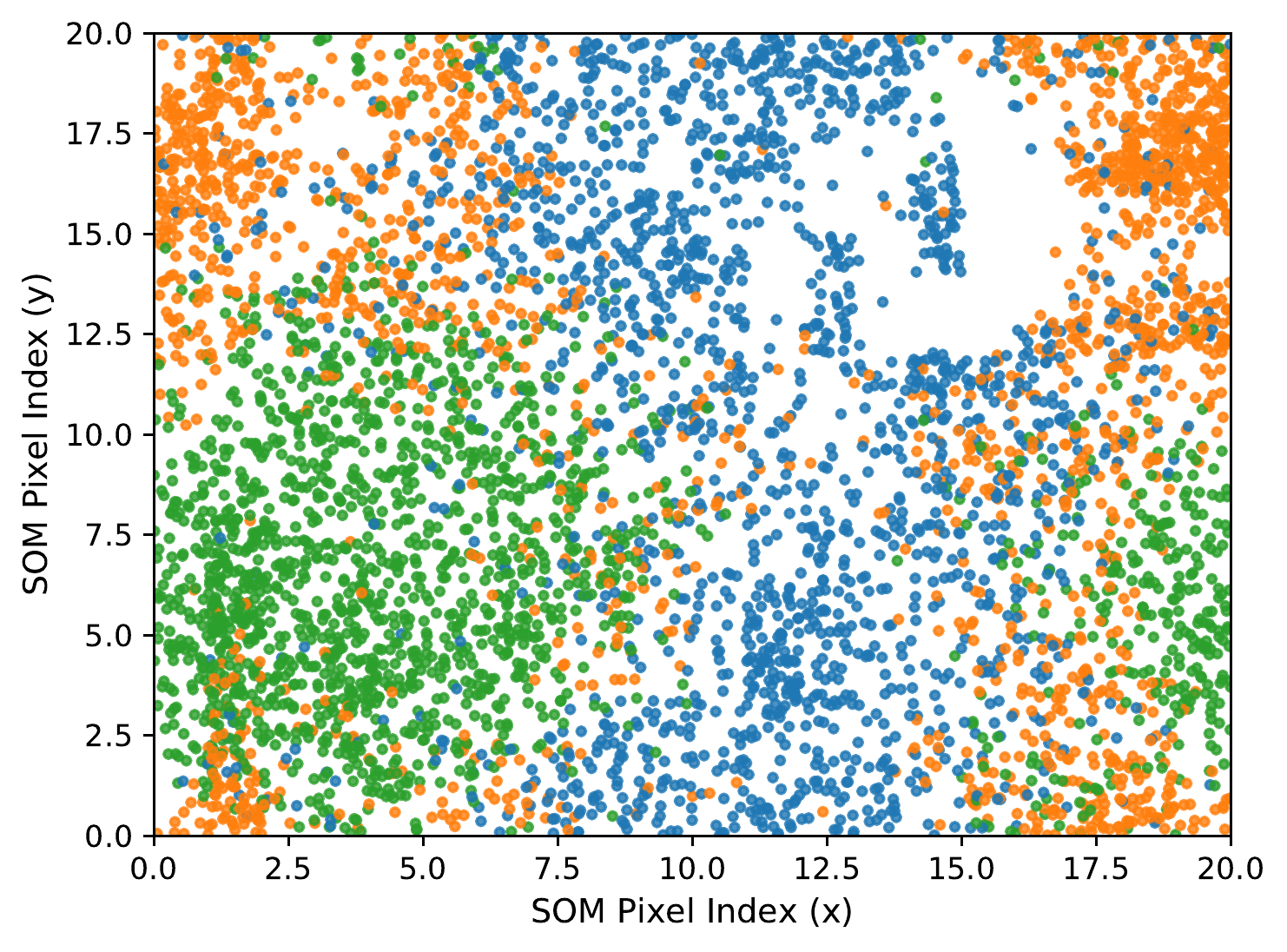}}
\caption{SOM pixel locations of labelled training set lightcurves, showing strong clustering. Green=planet, orange = astrophysical FP, blue=non-astrophysical FP. A random jitter of between -0.5 and 0.5 pixels has been added in both axes for clarity.}
\label{figsom}
\end{figure}

We consider several machine learning algorithms in Section \ref{sectModels}. Some algorithms are unlikely to deal well with direct lightcurve data, as it would dominate the feature space. For these we create a summary statistic for the lightcurves following the self-organising-map (SOM) method of \citet{Armstrong:2017cp}, applying our lightcurves to their publicly available \Kepler SOM. We create a further SOM statistic using the same methodology but with a SOM trained on our own dataset, to encourage discrimination of non-astrophysical FPs which weren't studied in \citet{Armstrong:2017cp}. The resulting SOM is shown in Figure \ref{figsom}. These SOM statistics are a form of dimensionality reduction, reducing the lightcurve shape into a single statistic.

For a given algorithm, either the two SOM statistics are appended to the TCE table feature set, or the `local' view lightcurve with 201 bin values is appended. As such we have two data representations, `Feature+SOM' and `Feature+LC'. The used features, and which models they apply to, are detailed in Table \ref{tabfeatures}.

\subsection{Minimally useful attributes}
\label{sectfeatimp}
It is desirable to reduce the feature space to the minimum useful set, so as to simplify the resulting model and reduce the proportion of non-informative features passed to the models. We drop columns from the TCE table using a number of criteria. Initially metadata associated with the table is dropped, including delivery name and \Kepler identifier. Columns associated with the error on another column are dropped. Columns associated with a trapezoid fit to the lightcurves are dropped in favour of the actual planet model fit also performed. We drop most centroid information, limiting the models to one column providing the angular offset between event centroids and the KIC position, finding that this performed better than differential measures. Columns related to the autovetter \citep{McCauliff:2015fb} are dropped, along with limb darkening coefficients, and the planet albedo and implied temperature are dropped in favour of their associated statistics which better represent the relevant information for planet validation. We further experimented with removing the remaining features in order to create a minimal set, finding that the results in fact marginally improved when we reduced the data table to the thirty eight features detailed in Table \ref{tabfeatures}, in addition to the SOM features or the local view lightcurve.



\begin{table*}
\caption{Data features. GPC=Gaussian Process Classifier, RF=Random Forest, ET=Extra Trees, MLP=Multilayer Perceptron.}
\label{tabfeatures}
\begin{tabular}{llrr}
\hline
\hline
Name & Description & In GPC & In RF/ET/MLP \\
\hline
tce\_period &  Orbital period of the TCE &  x & x \\
tce\_time0bk &  Centre time of the first detected transit in BJD &  x & x \\
tce\_ror & Planet radius dived by the stellar radius &  x & x \\
tce\_dor & Planet-star distance at mid-transit divided by the stellar radius &  x & x \\
tce\_duration & Duration of the candidate transit (hours) &  x & x \\
tce\_ingress & Ingress duration (hours) &  x & x \\
tce\_depth & Transit depth (ppm) &  x & x \\
tce\_model\_snr & Transit depth normalised by the mean flux uncertainty in transit &  x & x \\
tce\_robstat & A measure of depth variations across all transits &  x & x \\
tce\_prad & Implied planet radius &  x & x \\
wst\_robstat & As tce\_robstat for the most significant secondary transit &  x & x \\
wst\_depth & Fitted depth of the most significant secondary transit &  x & x \\
tce\_mesmedian & See \citet{Twicken:2018ca} &  x & x \\
tce\_mesmad & See \citet{Twicken:2018ca} &  x & x \\
tce\_maxmes & MES statistic of most significant secondary transit &  x & x \\
tce\_minmes & MES statistic of least significant secondary transit &  x & x \\
tce\_maxmesd & Phase in days of most significant secondary transit &  x & x \\
tce\_minmesd & Phase in days of least significant secondary transit &  x & x \\
tce\_max\_sngle\_ev & Maximum single event statistic &  x & x \\
tce\_max\_mult\_ev & Maximum multiple event statistic (MES) &  x & x \\
tce\_bin\_oedp\_stat & Odd-Even depth comparison statistic &  x & x \\
tce\_rmesmad & Ratio of MES to median average deviation (MAD) MES &  x & x \\
tce\_rsnrmes & Ratio of signal-to-noise ratio to MES &  x & x \\
tce\_rminmes & Ratio of minimum MES to MES &  x & x \\
tce\_tce\_albedostat & Significance of geometric albedo derived from secondary&  x & x \\
tce\_ptemp\_stat & Significance of effective temperature derived from secondary &  x & x \\
boot\_fap & Bootstrap false alarm probability &  x & x \\
tce\_cap\_stat & Ghost core aperture statistic &  x & x \\
tce\_hap\_stat & Ghost halo aperture statistic &  x & x \\
tce\_dikco\_msky & Angular offset between event centroids from KIC position &  x & x \\
max\_ephem\_corr & Maximum ephemeris correlation &  x & x \\
Kepler & Kepler magnitude &  x & x \\
Teff & Host stellar temperature &  x & x \\
Radius & Host stellar radius from \citet{Collaboration:2018io} &  x & x \\
tce\_model\_redchisq & Transit fit model reduced $\chi^2$ &  x & x \\
tce\_chisq1dof1 & See \citet{Tenenbaum:2013el} and \citet{Seader:2013fd} &  x & x \\
tce\_chisq1dof2 & See \citet{Tenenbaum:2013el} and \citet{Seader:2013fd} &  x & x \\
tce\_chisqgofdofrat & See \citet{Seader:2015ca} &  x & x \\
somstat & SOM statistic using new SOM trained on this data &   & x \\
a17stat & SOM statistic using SOM of \citet{Armstrong:2017cp} &   & x \\
Local View lightcurve & 201 bin local view of the transit lightcurve & x  &  \\
\hline
\end{tabular}
\end{table*}

\subsection{Data Scaling}
\begin{figure}
\label{figtsetdists}
\resizebox{\hsize}{!}{\includegraphics{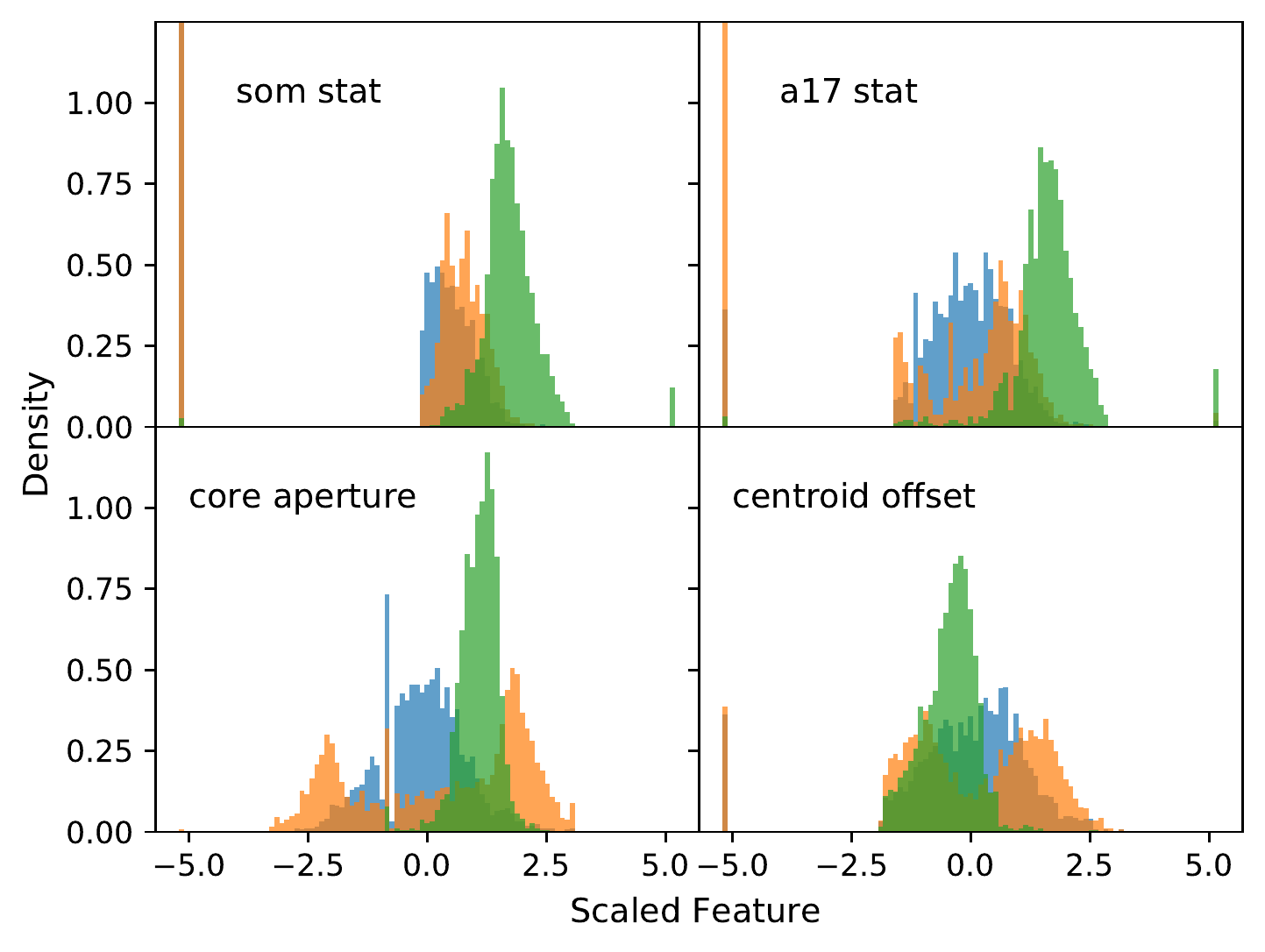}}
\caption{Training set distributions of the most important four features after scaling. Confirmed planets are in green, astrophysical FPs in orange, and non-astrophysical FPs in blue. The single value peaks occur due to large numbers of TCEs having identical values for a feature. The vertical axis cuts off some of the distribution in the top two panels to better show the overall distributions.}
\end{figure}

Many machine learning algorithms perform better when the input data is scaled. As such we scale each of our inputs to follow a normal distribution with a mean of zero and variance of unity for each feature. The only exceptions are the `local' view lightcurve values, which are already scaled. The most important four feature distributions as measured by the optimised random forest classifier (RFC) from Section \ref{sectModels} are plotted in Figure \ref{figtsetdists} after scaling.

\subsection{Training Set Dispositions}
Information on the disposition of each TCE is extracted from the DR25 ordinary and supplementary KOI tables (hereafter koi-base and koi-supp respectively). koi-base is the KOI table derived exclusively from DR25, whereas koi-supp contains a `best-knowledge' disposition for each KOI. We build our confirmed planet training set by taking objects labelled as confirmed in the koi-supp table (column `koi\_disposition'), which are in the koi-base table and not labelled as FPs or indeterminate in either \citet{Santerne:2016fz} or \citet{Burke:2019js}. This set includes previously validated planets. We remove a small number of apparently confirmed planets where the \Kepler data has shown them to be FPs, based on the `koi\_pdisposition' column. We use koi-supp to give the most accurate dispositions for individual objects, prioritising training set label accuracy over uniformly processed dispositions. This leaves 2274 TCEs labelled as confirmed planets.

\begin{figure}
\label{figtsetRPdists}
\resizebox{\hsize}{!}{\includegraphics{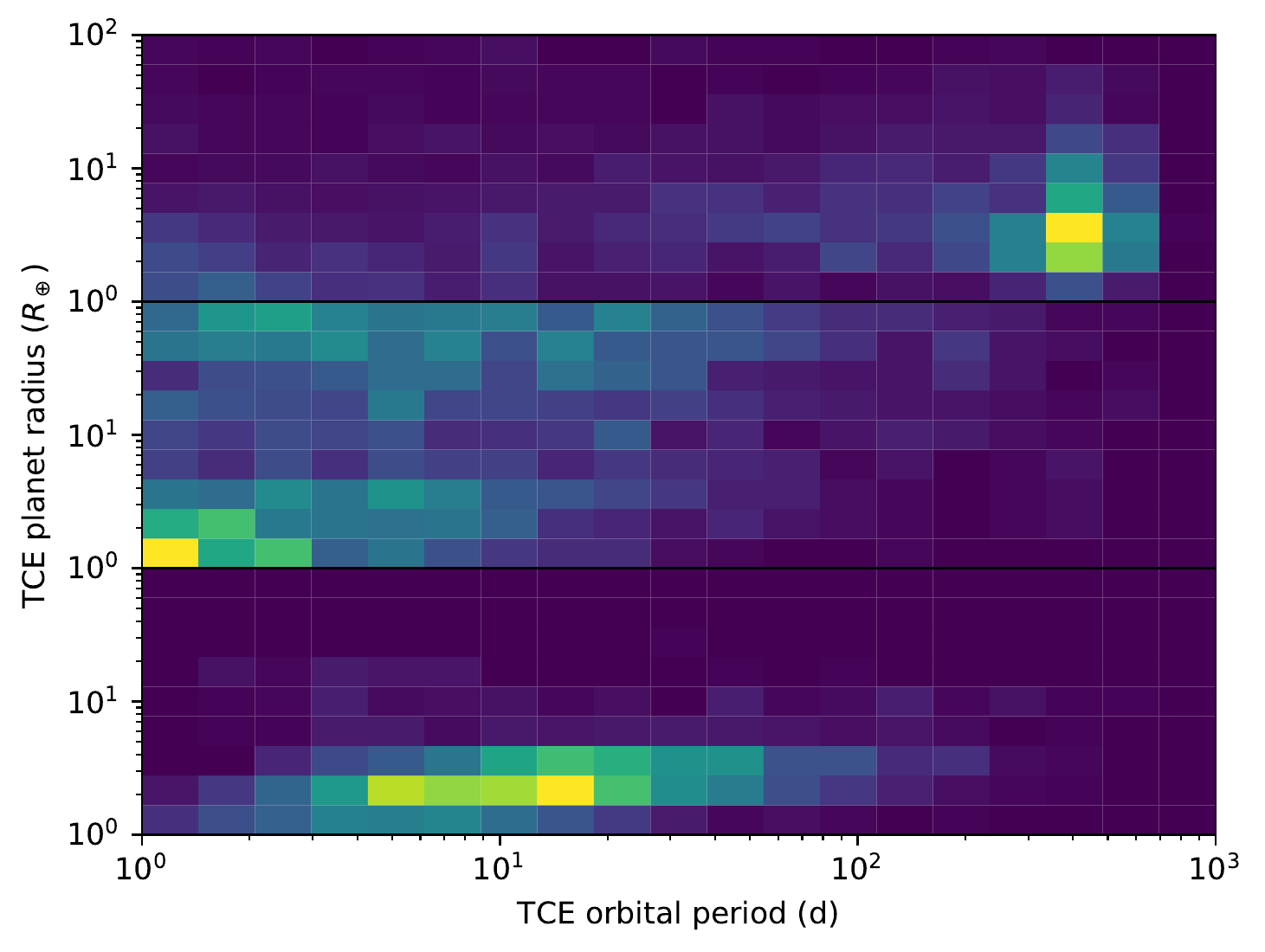}}
\caption{Training set planet radius and period distributions. Top: Non-astrophysical FPs. Middle: Astrophysical FPs. Bottom: planets. All distributions are normalised to show probability density. Training set members with apparent planet radii larger than 100$R_\oplus$ are not plotted for clarity (all are FPs).}
\end{figure}

We build two FP sets, one each for astrophysical and non-astrophysical FPs. The astrophysical FP set contains all KOIs labelled false positive in the koi-supp table (column `koi\_pdisposition'), which are in the koi-base table, where there is not a flag raised indicating a non-transiting-like signal, and supplemented by all false positives in \citet{Santerne:2016fz}. The non-astrophysical FP set contains KOIs where a flag was raised indicating a non-transiting-like signal, supplemented by 2200 randomly drawn TCEs which were not upgraded to KOIs. By utilising these random TCEs we are implicitly assuming that the TCEs which were not made KOIs are in the majority FPs, which is born out by our results (Section \ref{sectResnonKOI}). The astrophysical FP set then has 3100 TCEs, and the non-astrophysical FP set has 2959 TCEs. The planet radius and period distributions of the three sets are shown in Figure \ref{figtsetRPdists}.

We combine the two FP sets going forwards, leaving a FP set with approximately double the number of the confirmed planet set. This imbalance will be corrected implicitly by some of our models, but in cases where it isn't, or in case the correction is not effective, this overabundance of FPs ensures that any bias in the models prefers FP classifications. We do not include additional TCEs to avoid unbalancing the training sets further, which can impact model performance.

We split our data into a training set (80\%, 6663 TCEs), a validation set for model selection and optimisation (10\%, 834 TCEs) and a test set for final analysis of model performance (10\%, 836 TCEs), in each case maintaining the proportions of planets to FPs. TCEs with no disposition form the 'unknown' set (24201 TCEs).

\subsection{Training Set Scenario Distributions}
\label{secttsetdists}
The algorithms we are building fundamentally aim to derive the probability that a given input is a member of one of the given training sets. As such the membership, information in, and distributions of the training sets are crucially important. The overall proportion of FPs relative to planets is deliberately left to be incorporated as prior information. We could attempt to include it by changing the relative numbers within the confirmed planet and FP datasets, but the number of objects in a training set is not trivially related to the output probability for most machine learning algorithms.

Another consideration is the relative distributions of object parameters within each of the planet and FP datasets. This is where the effect of, for example, planet radius on the likelihood of a given TCE being a FP will appear. By taking the confirmed and FP classifications of the koi-supp table as our input, we are implicitly building in any biases present in that table into our algorithm. We note that the table distribution is in part the real distribution of planets and FPs detected by the \Kepler satellite and the detection algorithms which created the TCE and KOI lists. Incorporating that distribution is in fact desirable, given we are studying candidates found using the same process, and in that sense the \Kepler set of planets and FPs is the ideal distribution to use.

The distribution of \Kepler detected planets and FPs we use will however be biased by the methods used to label KOIs as planets and FPs. In particular the majority of confirmed planets and many FPs labelled in the KOI list have been validated by the \texttt{vespa} algorithm (\mytilde 50\% of the known KOI planets), and as such biases in that algorithm may be present in our results. We compare our results to the \texttt{vespa} designations in Section \ref{sectVespaComp}, showing they disagree in many cases despite this reliance on \texttt{vespa} designations. The reliance on past classification of objects as planet or FP is a weakness of our method which we aim to improve in future work, using simulated candidates from each scenario.

A further point is the balance of astrophysical to non-astrophysical FPs in the training set. We can estimate what this should be using the ratio of KOIs to TCEs, where KOIs are \mytilde 30\% of the TCE list, under the assumption that the majority of non-KOI TCEs are non-astrophysical FPs. We use a 50\% ratio in our training set, which effectively increases the weighting for the astrophysical FPs. This ratio improves the representation of astrophysical FPs, which is desirable given that non-astrophysical FPs are easy to distinguish given a high enough signal-to-noise. We impose a MES cut of 10.5 as recommended by \citet{Burke:2019js} before validating any candidate to remove the possibility of low signal-to-noise instrumental FPs complicating our results.




\section{Model Selection and Optimisation}
\label{sectModels}
Many machine learning methods are available, with a range of complexity and properties. We perform empirical model selection using the two input data sets. For the Feature+SOM set, we implement eight models with a range of parameters, testing a total of 822 combinations, using the \texttt{scikit-learn} \texttt{python} module \citep{Pedregosa:2011cx}. The best parameters for each algorithm were selected by comparing scores on the validation set. The trialled model parameters are shown in Table \ref{taboptparams}, with the best found parameters highlighted. The final performances of each model are given in Table \ref{tabscores}, with and without probability calibration which is described in Section \ref{sectCal}, and are measured using the log-loss metric \citep[see e.g.][]{Malz:2019ib} calculated on the test set. The log-loss is given by
\begin{equation}
L_{log} = -\frac{1}{N}\sum_{i=0}^{N-1}(y_i\log(p_i)+(1-y_i)\log(1-p_i))
\end{equation}
\noindent where $y_i$ is the true class label of candidate $i$ and $p_i$ is its output classifier score, and $N$ is the number of test samples.

The utilised models are described in Section \ref{sectFramework}, but readers interested in the other models are referred to the \texttt{scikit-learn} documentation and references therein \citep{Pedregosa:2011cx}. 

We found that while most tested models were highly successful, the best performance after calibration was shown by a RFC. It is interesting to see the relative success of even very simple models such as Linear Discriminant Analysis (LDA), implying the underlying decision space is not overly complex. The overall success of the models is not unexpected, as we are providing the classifiers with very similar information as was often used to classify candidates as planets or FPs in the first place, and in the case of \texttt{vespa} validated candidates, we are adding more detailed lightcurve information. We proceed with the RFC as a versatile robust algorithm, supplementing the results with classifications from the next two most successful models, Extra Trees (ET) and Multi-layer Perceptron (MLP), to guard against overfitting by any one model.

\begin{figure}
\label{figARDscales}
\resizebox{\hsize}{!}{\includegraphics{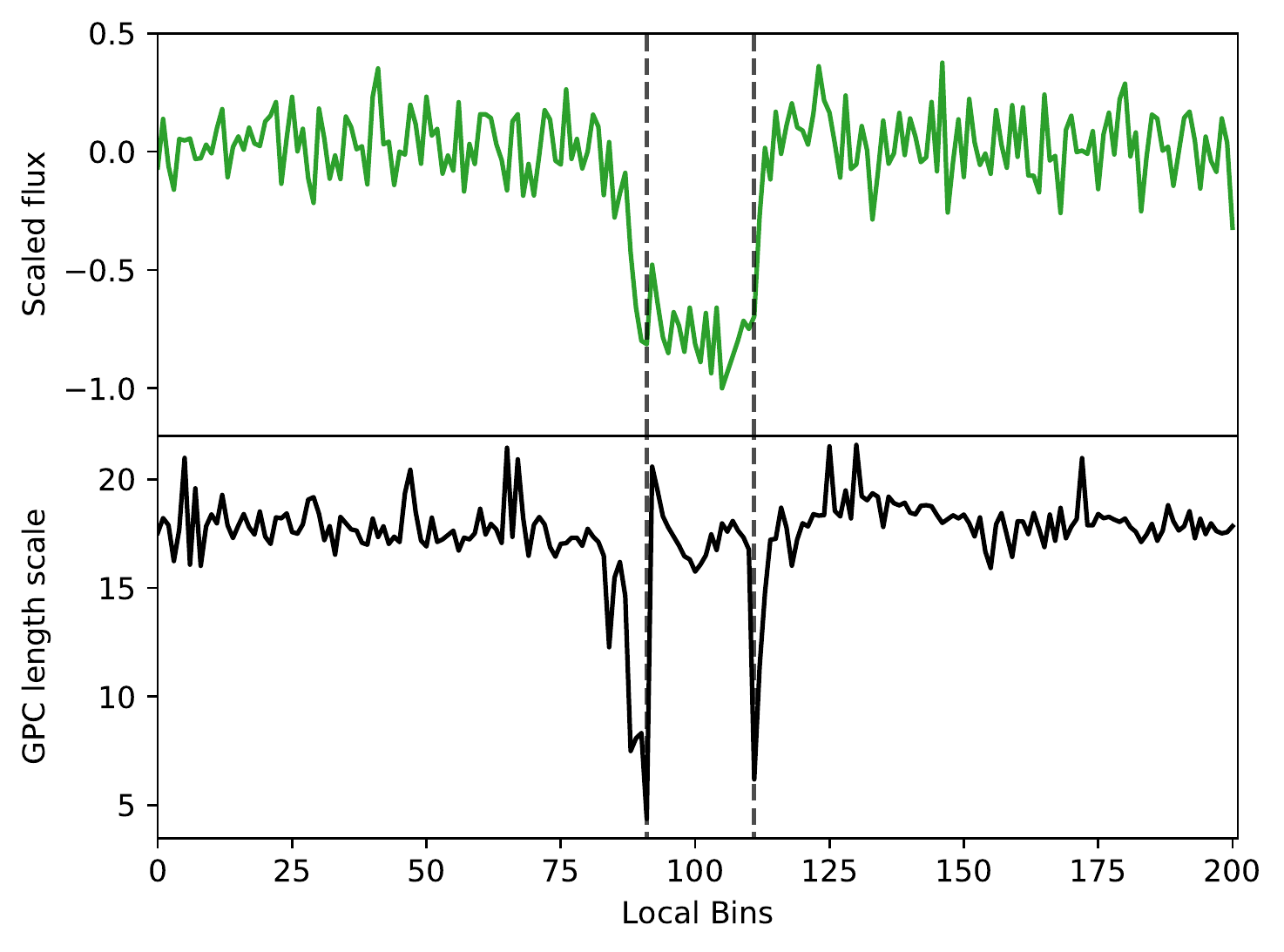}}
\caption{Top: Local view of a planet lightcurve. Bottom: GPC automatic relevance determination (ARD) lengthscales for each of the input bins in the local view. Low lengthscales can be seen at ingress and egress, demonstrating that the GPC has learned to prioritise those regions of the lightcurve when making classifications.}
\end{figure}

For the Feature+LC input data, we utilise a gaussian process classifier (GPC) to provide an independent and naturally probabilistic method for comparison and to guard against overconfidence in model classifications. We implement the GPC using \texttt{gpflow}. The GPC is optimised varying the selected kernel function, and final performance is shown in Table \ref{tabscores}. Additionally we trial the GPC using variations of the input data - with the Feature+SOM data, lightcurve and a subset of features (Features+LC-light) and with the full Feature+LC dataset. We find the results are not strongly dependent on input dataset, and hence use the Feature+LC dataset to provide a difference to the other models. Figure \ref{figARDscales} shows the GPC adapting to the input transit data. The underlying theory of a GPC was summarised in Section \ref{sectGPC}.

\begin{table}
\caption{Trialled model parameters. All combinations listed were tested. Best parameters as found on the validation set are in bold.}
\label{taboptparams}
\begin{tabular}{lr}
\hline
\hline
\textbf{Model} & \\
Parameter & Options  \\
\hline
\textbf{GPC} & \\
n\_inducing\_points & [16,\textbf{32},64,128]\\
likelihood & [\textbf{bernoulli}]\\
kernel & [rbf,linear,\textbf{matern32},matern52,\\
   &  polynomial (order 2 and 3)] \\
ARD weights & [\textbf{True}, False] \\

\textbf{RFC} & \\
n\_estimators & [300,500,1000,\textbf{2000}] \\
max\_features & [5,\textbf{6},7] \\
min\_samples\_split & [2,3,4,\textbf{5}] \\
max\_depth & [\textbf{None},5,10,20] \\
class\_weight & [\textbf{balanced}] \\

\textbf{Extra Trees} & \\
n\_estimators & [300,500,1000,\textbf{2000}] \\
max\_features & [5,6,\textbf{7}] \\
min\_samples\_split & [2,3,4,\textbf{5}] \\
max\_depth & [None,5,10,\textbf{20}] \\
class\_weight & [\textbf{balanced}] \\

\textbf{Multilayer Perceptron} & \\
solver & [\textbf{adam},sgd] \\
alpha & [1,1e-1,1e-2,\textbf{1e-3},1e-4,1e-5] \\
hidden\_layer\_sizes & [(10,),\textbf{(15,)},(20,),(5,5),(5,10)] \\
learning\_rate & [constant,invscaling,\textbf{adaptive}] \\
early\_stopping & [True,\textbf{False}] \\
max\_iter & [\textbf{2000}] \\

\textbf{Decision Tree} & \\
max\_depth & [10,\textbf{20},30] \\
class\_weight & [\textbf{balanced}] \\

\textbf{Logistic} & \\
penalty & [\textbf{l2}] \\
class\_weight & [\textbf{balanced}] \\

\textbf{QDA} & \\
priors & [\textbf{None}] \\


\textbf{K-NN} & \\
n\_neighbours & [3,5,7,\textbf{9}]\\
metric & [minkowski,\textbf{euclidean},manhattan]\\
weights & [\textbf{uniform},distance]\\

\textbf{LDA} & \\
priors & [\textbf{None}] \\


\hline
\end{tabular}
\end{table}


\begin{table*}
\caption{Best model performance on test set, ranked by calibrated log-loss. The GPC does not require external calibration.}
\label{tabscores}
\begin{tabular}{lrrrrr}
\hline
\hline
Model & AUC & Precision & Recall & Log-loss & Calibrated Log-loss  \\
\hline
Gaussian Process Classifier & 0.999 & 0.984 & 0.995 & 0.54 & ---\\
Random Forest & 0.999 & 0.981 & 0.997 &  0.58  & 0.54 \\
Extra Trees  & 0.999 & 0.985 & 0.992  & 0.58 & 0.58 \\
Multilayer Perceptron  & 0.997 & 0.982 &  0.985 & 0.83  & 0.66 \\
K-Nearest Neighbours  & 0.997 & 0.995 &  0.972 & 0.83 & 0.66 \\
Decision Tree  & 0.958 & 0.979 & 0.984  & 0.95  & 0.74 \\
Logistic Regression  & 0.997 & 0.988 & 0.967  & 1.12 & 1.03  \\
Quadratic Discriminant Analysis   & 0.989 & 0.983 & 0.970 & 1.16 & 1.20 \\
Linear Discriminant Analysis  & 0.993 & 0.982 &  0.965 & 1.32 & 1.36\\
\hline
\end{tabular}
\end{table*}

\section{Planet Validation}
\label{sectVal}

\subsection{Probability Calibration}
\label{sectCal}
\begin{figure}
\label{figuncal}
\resizebox{\hsize}{!}{\includegraphics{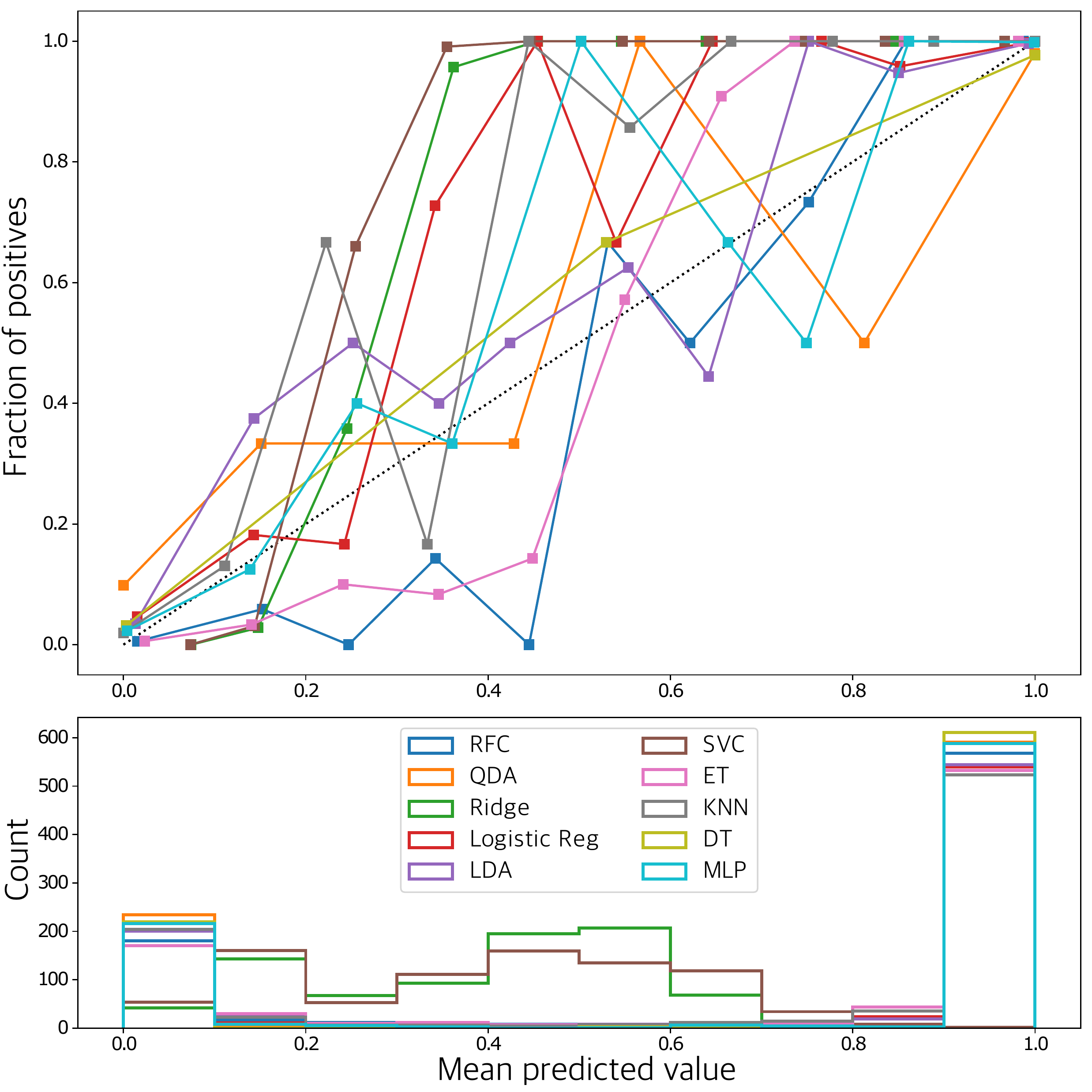}}
\caption{Top: Calibration curve for the uncalibrated non-GP classifiers. The black dashed line represents perfect calibration. Bottom: Histogram of classifications showing the number of candidates falling in each bin for each classifier.}
\end{figure}

Although the GPC naturally produces probabilities as output $p(s=1\vert\ \mathbf{x}^{*})$, the other classifiers are inherently non-probabilistic models and need to have their ad-hoc probabilities calibrated \citep{zadrozny2001obtaining,zadrozny2002transforming,niculescu2005predicting}. Classifier probability calibration is typically performed by plotting the `calibration curve', the fraction of class members as a function of classifier output. The uncalibrated curve is shown in Figure \ref{figuncal}, which highlights a counterintuitive issue; the better a classifier performs, the harder it can be to calibrate, due to a lack of objects being assigned intermediate values. Given our focus is to validate planets, we focus on accurate and precise calibration at the extreme ends, where $p(s=1\vert\ \mathbf{x}^{*})<0.01$ or $p(s=1\vert\ \mathbf{x}^{*})>0.99$.

To statistically validate a candidate as a planet, the commonly accepted threshold is $p(s=1\vert\mathbf{x}^{*})>0.99$ \citep{Morton:2016ka}. Measuring probabilities to this level requires the precision of our calibration is also at least 1\% or better. We use the \textit{isotonic regression} calibration technique \citep{zadrozny2001obtaining,zadrozny2002transforming}, which calibrates by counting samples in bins of given classifier scores. To measure the fraction of true planets in the $p(s=1\vert\mathbf{x}^{*})>0.99$ bin we therefore require at least $N=10000$ test planets to reduce the Poisson counting error $\sqrt{N}/N$ below 1\%. Given the size of our training set additional test inputs are required for calibration. 

To allow calibration at this precision, we synthesise additional examples of planets and FPs from our training set, by interpolating between members of each class. The process is only performed for the Feature+SOM dataset, as the GPC does not need calibrating. We select a training set member at random, and then select another member of the same class which is within the 20th percentile of all the member-to-member distances within that class. Distances are calculated by considering the euclidean distance between the values of each column for two class members. Restricting the distances in this way allows for non-trivial class boundaries in the parameter space. A new synthetic class member is then produced by interpolating between the two selected real inputs. We generate 10000 each of planets and FPs from the training set. These synthetic datasets are used only for calibration, not to train the classifiers.

It is important to note that by interpolating, we have essentially weakened the effect of outliers in the training data, at least for the calibration step. For this and other reasons, candidates which are outliers to our training set will not get valid classifications, and should be ignored. We describe our process for flagging outliers in Section \ref{sectOutliers}. Interpolation also means that while we can attain the desired precision, the accuracy of the calibration may still be subject to systematic biases in the training set, which were discussed in Section \ref{secttsetdists}.



\begin{figure}
\label{figcal}
\resizebox{\hsize}{!}{\includegraphics{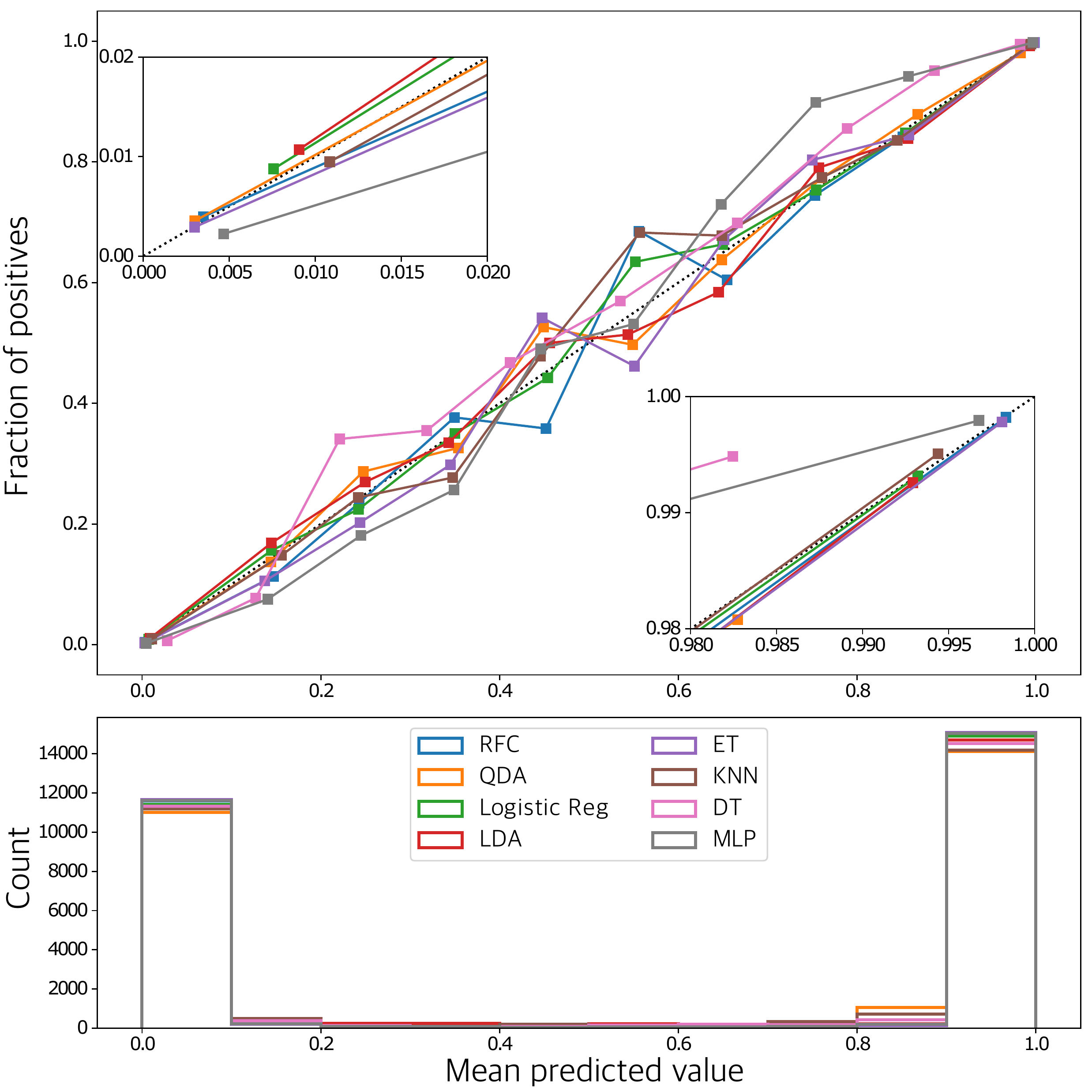}}
\caption{Top: Calibration curve for the calibrated non-GP classifiers. The black dashed line represents perfect calibration. The two insets show zoomed plots of the low and high ends of the curve. Bottom: Histogram of classifications showing the number of candidates falling in each bin for each classifier. Synthetic training set members are included in this plot.}
\end{figure}

\subsection{Classifier training}

The GPC was trained using the training data with no calibration. We used the \texttt{gpflow} \citep{deGMatthews:2017tm} \texttt{python} extension to \texttt{tensorflow} \citep{Abadi:2016vn}, running on an NVidia GeForce GTX Titan XP GPU. On this architecture the GPC takes less than one minute to train, and seconds to classify new candidates.

For the other classifiers, training and calibration was performed on a 2017 generation iMac with four 4.2GHz Intel i7 processors. Training each model takes a few minutes, with classification of new objects possible in seconds. To create calibrated versions of the other classifiers, we employ a cross validation strategy to ensure that the training data can be used for training and calibration. The training set and synthetic dataset are split into 10 folds, and on each iteration a classifier is trained on 90\% of the training data, then calibrated on the remaining 10\% of training data plus 10\% of the synthetic data. The process is repeated for each fold to create ten separate classifiers, with the classifier results averaged to produce final classifications.

The above steps suffice to give results on the validation, test and unknown datasets.  We also aim to classify the training dataset independently, as a sanity check and to confirm previous validations. To get results for the training dataset we introduce a further layer of cross-validation, with 20 folds. For the GPC this is the only cross validation, where the GPC is trained on 95\% of the training data to give a result for the remaining 5\%, and the process repeated to classify the whole training set. For the other classifiers we separate 5\% of the training data before performing the above training and calibration steps using the remaining 95\%, and repeat.


\subsection{Positional Probabilities}
\label{sectPosProbs}
Part of the prior probability for an object to be a planet or FP, $P(s | \text{I})$, is the probability that the signal arises from the target star, a known blended star in the aperture, or an unresolved background star. We derive these values using the positional probabilities calculated in \citet{Bryson:tz}, which provide the probability that the signal arises from the target star $P_\textrm{target}$, the probability arises from a known secondary source $P_\textrm{secondsource}$, typically another KIC target or a star in the \Kepler UKIRT survey\footnote{https://keplerscience.arc.nasa.gov/community-products.html}, and the probability the signal arises from an unresolved background star $P_\textrm{background}$. \citet{Bryson:tz} also considered a small number of sources detected through high resolution imaging; we ignore these and instead take the most up to date results from the robo-AO survey from \citet{Ziegler:2018ei}. The given positional probabilities have an associated score representing the quality of the determination; where this is below the accepted threshold of 0.3 we continue with the a priori values given by \citet{Bryson:tz}, but do not validate planets where this occurs.

The calculation in \citet{Bryson:tz} was performed without Gaia DR2 \citep{Collaboration:2018io} information, and so we update the positional probabilities using the new available information. We first search the Gaia DR2 database for any detected sources within 25\arcsec of each TCE host star. We chose 25\arcsec as this is the limit considered for contaminating background sources in \citet{Bryson:tz}. Gaia sources which are in the KIC \citep[identified in][]{Berger:2018il}, in the Kepler UKIRT survey or in the new robo-AO companion source list are discarded as these were either accounted for in \citet{Bryson:tz} or are considered separately in the case of robo-AO.

We then check for each TCE whether any new Gaia or robo-AO sources are bright enough to cause the observed signal, conservatively assuming a total eclipse of the blended background source. If there are such sources, we flag the TCE in our results and adjust the probability of a background source causing the signal $P_\textrm{background}$ to account for the extra source, by increasing the local density of unresolved background stars appropriately and normalising the set of positional probabilities given the new value of $P_\textrm{background}$. It would be ideal to treat the Gaia source as a known second source, but without access to the centroid ellipses for each candidate we cannot make that calculation. We do not validate TCEs with a flag raised for a detected Gaia or robo-AO companion, although we still provide results in Section \ref{sectResults}.

\subsection{Prior Probabilities}
\label{sectPriors}
To satisfy Equation \ref{eqnmain2} we need the prior probability of a given candidate being a planet or FP, $P(s | \text{I})$, independently of the candidate parameters.  This prior probability for the planet scenario is given by

\begin{equation}
P(s=1 | \text{I})= P_\textrm{target} f_\textrm{planet} f_\textrm{transit}
\end{equation}

\noindent where $P_\textrm{target}$ is the probability of a signal arising from the host star and was calculated in Section \ref{sectPosProbs}, $f_\textrm{planet}$ is the probability of a randomly chosen star hosting a planet that \Kepler could detect and $f_\textrm{transit}$ represents the probability of that planet transiting, on average over the \Kepler candidate distribution. The product $f_\textrm{planet} f_\textrm{transit}$ represents the probability that a randomly chosen \Kepler target star hosts a planet which could have been detected by the \Kepler pipeline. We derive the product $f_\textrm{planet} f_\textrm{transit}$ using the occurrence rates calculated by \citet{Hsu:2018ec}, for planets with periods less than 320d and radii between 2 and 12 $R_\oplus$. We take each occurrence rate bin in their paper, calculate the eclipse probability for a planet in the centre of the bin to transit a solar host star, and sum the resulting probabilities to get a final product $f_\textrm{planet} f_\textrm{transit} = 0.0308$. The effect of specific planet radius, period and host star is included in the classification models.



We consider several FP scenarios and sum their probabilities to give the overall prior for FPs. We take

\begin{equation}
\begin{split}
P(s = 0 | \text{I}) =& P(\textrm{FP-EB})+ P(\textrm{FP-HEB}) \\ & + P(\textrm{FP-HTP}) \\ & + P(\textrm{FPresolved}) \\ & + P(\textrm{FP-BEB}) + P(\textrm{FP-BTP})\\ & + P(\textrm{FPnon-astro})
\end{split}
\end{equation}

\noindent where $P(\textrm{FP-EB})$ is the prior for an eclipsing binary on the target star, $P(\textrm{FP-HEB})$ is the prior for a hierarchical eclipsing binary, i.e. a triple system where the target star has an eclipsing binary companion causing the signal, and $P(\textrm{FP-HTP})$ is the prior for a hierarchical transiting planet, i.e. a planet transiting the fainter companion in a binary system. $P(\textrm{FPresolved})$ is the prior for a transiting planet, eclipsing binary or hierarchical eclipsing binary on a resolved non-target star. We disregard hierarchical transiting planets on second known sources as contributing insignificantly towards the FP probability. $P(\textrm{FP-BEB})$ and $P(\textrm{FP-BTP})$ are the priors for an eclipsing binary or a transiting planet on an unresolved background star. $P(\textrm{FPnon-astro})$ is the prior for an instrumental or otherwise non-astrophysical source of the signal. We do not consider planets transiting the target star to be FPs even in the case where other stars, bound or otherwise, are diluting the signal. In our methodology these priors are independent of the actual orbital period of the contaminating binary, and so TCE FPs where the FP is an eclipsing binary with half the actual binary orbital period, as seen in \citet[e.g.][]{Morton:2016ka}, are covered by the same priors.

For the scenario specific priors,

\begin{equation}
P(\textrm{FP-EB}) = P_\textrm{target} f_\textrm{close-binary} f_\textrm{eclipse}
\end{equation}

\begin{equation}
P(\textrm{FP-HEB}) = P_\textrm{target} f_\textrm{close-triple} f_\textrm{eclipse}
\end{equation}

\begin{equation}
P(\textrm{FP-HTP}) = P_\textrm{target} f_\textrm{binary} f_\textrm{planet} f_\textrm{transit}
\end{equation}

\begin{multline}
P(\textrm{FPresolved}) = P_\textrm{secondsource} (f_\textrm{close-binary} f_\textrm{eclipse} +\\ f_\textrm{close-triple} f_\textrm{eclipse} + f_\textrm{planet} f_\textrm{transit})
\end{multline}

\begin{equation}
P(\textrm{FP-BEB}) = P_\textrm{background}(f_\textrm{close-binary} f_\textrm{eclipse})
\end{equation}

\begin{equation}
P(\textrm{FP-BTP}) = P_\textrm{background} (f_\textrm{planet} f_\textrm{transit})
\end{equation}

where $P_\textrm{target}$, $P_\textrm{secondsource}$ and $P_\textrm{background}$ were derived in Section \ref{sectPosProbs}. We discuss each prior in turn.

\subsubsection{$P(\textrm{FP-EB})$}
To calculate $P(\textrm{FP-EB})$ we need the probability of a randomly chosen star being an eclipsing binary with an orbital period $P$ which \Kepler could detect. We calculate the product $f_\textrm{close-binary} f_\textrm{eclipse}$ using the results of \citet{Moe:2017eg}. We integrate their occurrence rate for companion stars to main sequence solar-like hosts as a function of $\log P$ (their equation 23) multiplied by the eclipse probability at that period for a solar host star. We consider companions with $\log P<2.5$ ($P<320$d) and mass ratio $q>0.1$, correcting from the $q>0.3$ equation using a factor of 1.3 as suggested. The integration gives $f_\textrm{close-binary} f_\textrm{eclipse} = 0.0048$, which is strikingly lower than the planet prior, primarily due to the much lower occurrence rate for close binaries. Ignoring eclipse probability, we find the frequency of solar-like stars with companions within 320d to be 0.055 from \citet{Moe:2017eg}. It is often stated that \mytilde 50\% of stars are in multiple systems, but this fraction is dominated by wide companions with orbital periods longer than 320d. This calculation implicitly assumes that any eclipsing binary in this period range with mass ratio greater than 0.1 would lead to a detectable eclipse in the \Kepler data.

\subsubsection{$P(\textrm{FP-HEB})$}
The probability that a star is a hierarchical eclipsing binary depends on the triple star fraction. In our context the product $f_\textrm{close-triple}f_\textrm{eclipse}$ is the probability for a star to be in a triple system, where the close binary component is in the background, has an orbital period short enough for \Kepler to detect, and eclipses. The statistics for triple systems of this type (A-(Ba,Bb)) are extremely poor \citep{Moe:2017eg} due to the difficulty of reliably detecting additional companions to already lower mass companion stars. If we assume that one of the B components is near solar-mass, then we can use the general close companion frequency, which is the same as $f_\textrm{close-binary} f_\textrm{eclipse}$, multiplied by an additional factor to account for an additional wider companion. We use the fraction of stars with any companion from \citet{Moe:2017eg}, which is $f_\textrm{multiple}=0.48$. As such we take $f_\textrm{close-triple}f_\textrm{eclipse} = f_\textrm{multiple}f_\textrm{close-binary} f_\textrm{eclipse}$. Again this calculation implicitly assumes that any such triple with mass ratios greater than 0.1 to the primary star would lead to a detectable eclipse in the \Kepler data.

\subsubsection{$P(\textrm{FP-HTP})$}
Unlike the stellar multiple cases it is unlikely that all background transiting planets would produce a detectable signal in the \Kepler data. Estimating the fraction that do is complex and would require an estimate of the transit depth distribution for the full set of background transiting planets. Instead we proceed with the assumption that all such planets would produce a detectable signal if in a binary system, but not in systems of higher order multiplicity. $f_\textrm{binary}=0.27$ from \citet{Moe:2017eg} for solar-like primary components, which is largely informed by \citet{Raghavan:2010gd}. $f_\textrm{planet} f_\textrm{transit} = 0.0308$ as calculated above. Note we do not include any effect of multiplicity on the planet occurrence rate.

All necessary components for the remaining priors have now been discussed, although we again note the implicit assumption that all scenarios could produce a detectable transit.

\subsubsection{$P(\textrm{FPnon-astro})$}
$P(\textrm{FPnon-astro})$ is difficult to calculate, and so we follow \citet{Morton:2016ka} in setting it to 5e-5. Recent work has suggested that the systematic false alarm rate is highly important when considering long period small planetary candidates \citep{Burke:2019js} and can be the most likely source of FPs for such candidates. The low prior rate for non-astrophysical FPs used here is justified because we apply a cut on the multiple event statistic (MES) of 10.5 as recommended by \citet{Burke:2019js}, allowing only significant candidates to be validated. At such an MES, the ratio of the systematic to planet prior is less than $10^{-3}$ \citep[][their Figure 3]{Burke:2019js}, which translates to a prior of order $10^{-5}$ when applied to our planet scenario prior.

\subsubsection{Prior information in the training set}
Note that the probability of the signal arising from the target star is included in our scenario prior as $P_\textrm{target}$. As some centroid information is included in the training data the classifiers may incorporate the probability of the signal arising from the target star internally. As such we are at risk of double counting this information in our posterior probabilities. We include positional probabilities in $P(s | \text{I})$ because the probabilities available from \citet{Bryson:tz} include information on nearby stars and their compatibility with the centroid ellipses derived for each TCE. This is more information than we can easily make available to the classifiers, and additionally improves interpretability by exposing the positional probabilities directly in the calculation. Removing centroid information from the classifiers would artificially reduce their performance. Including prior information on the target in both the classifiers and external prior is the conservative approach, because a significant centroid offset, or low target star positional probability, can only reduce the derived probability of a TCE being a planet.

\subsection{Outlier Testing}
\label{sectOutliers}
Our method is only valid for `inliers', candidates which are well represented by the training set and which are not rare or unusual. We perform two tests to flag outlier TCEs, using different methodologies for independence. 

The first considers outliers from the entire set of TCEs, to avoid mistakenly validating a candidate which is a unique case and hence might be misinterpreted. We implement the local outlier factor method \citep{Breunig:NVjM7mq0}, which measures the local density of an entry in the dataset with respect to its neighbours. The result is a factor which decreases as the local density drops. If that factor is particularly low, the entry is flagged as an outlier. We use a default threshold of -1.5 which labels 391 (1.2\%) of TCEs as outliers, 108 of which were KOIs. The local outlier factor is well suited to studying the whole dataset as it is an unsupervised method which requires no separate training set.

\begin{figure}
\label{figoutlierscores}
\resizebox{\hsize}{!}{\includegraphics{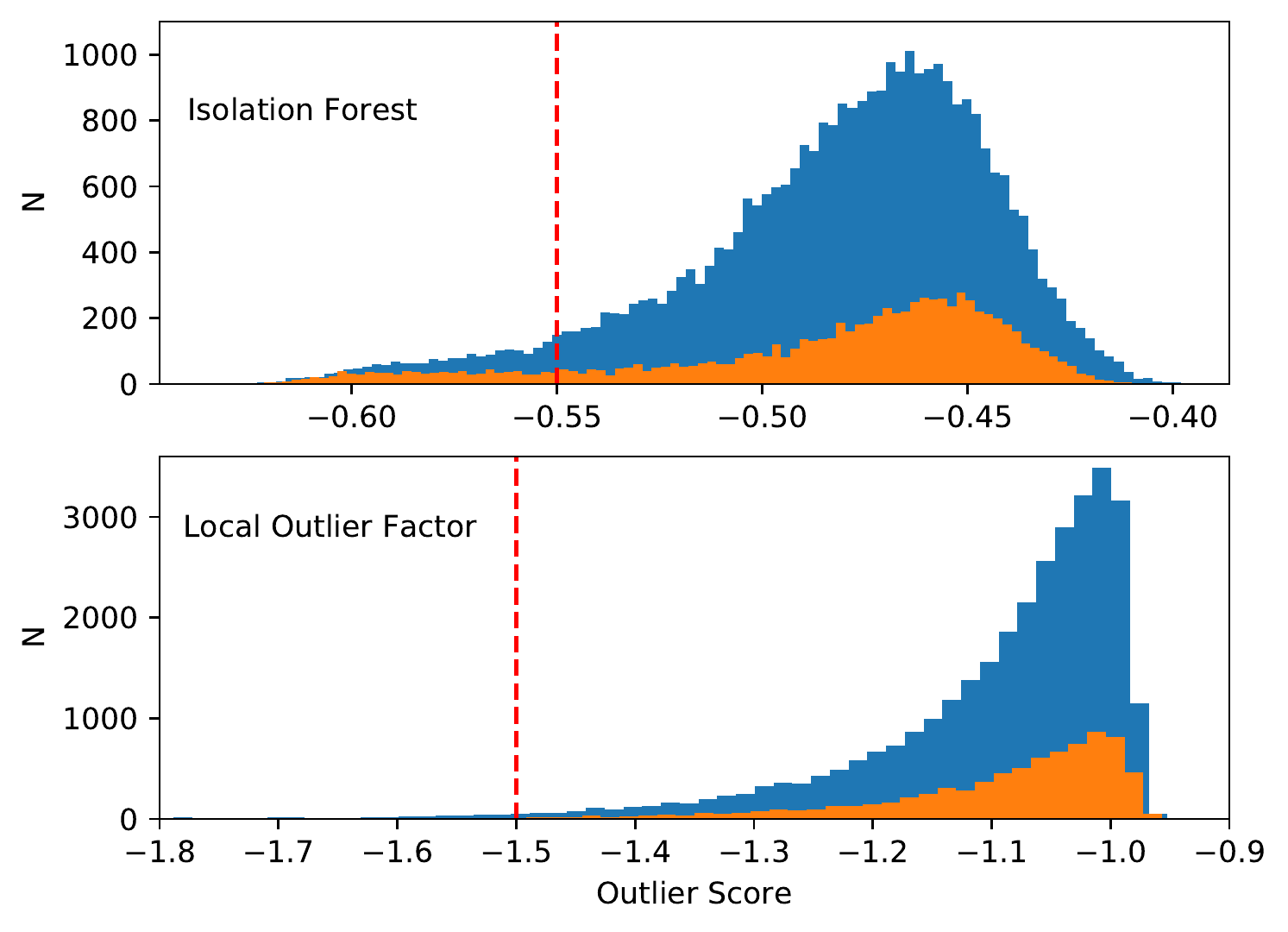}}
\caption{Top: Isolation Forest outlier score for all TCEs in blue and KOIs in orange. The red dashed line represents the threshold for outlier flagging. Bottom: As top for Local Outlier Factor. In both case outliers have more negative scores.}
\end{figure}

The second outlier detection method aims to find objects which are not well represented in the training set specifically. In this case we implement an isolation forest \citep{Liu:vni7ilVg} with 500 trees, which is trained on the training set then applied to the remaining data. Figure \ref{figoutlierscores} shows the distribution of scores produced from the isolation forest, where lower scores indicate outliers. The majority of candidates show a normal distribution, with a tail of more outlying candidates. We set our threshold as -0.55 based on this distribution, which flags 1979 (6.1\%) of TCEs as outliers, 932 of which were KOIs.

\subsection{External Flags}
Some information is only available for a small fraction of the sample, and hence is hard to include directly in the models. In these cases, we create external flags along with our model scores, and conservatively withhold validation from planets where a warning flag is raised. 

As described in Section \ref{sectPosProbs}, we flag TCEs where either Gaia DR2 or Robo-AO has detected a previously unresolved companion in the aperture bright enough to cause the observed TCE. The robo-AO flag supersedes the Gaia flag, in that if a source is seen in robo-AO we will not raise the Gaia flag for the same source. We also flag TCEs where the host star has been shown to be evolved in \citet{Berger:2018il} using the Gaia DR2 data, and include the \citet{Berger:2018il} binarity flag which indicates evidence for a binary companion from either the Gaia parallax or alternate high resolution imaging.

\subsection{Training Set Coverage}
\label{sectTsetCoverage}
It is crucial to be aware of the content of our training set: planet types or FP scenarios which are not represented will not be well distinguished by the models. The training set here is drawn from the real detected \Kepler distribution of planets and FPs, but potential biases exist for situations which are hard to disposition confidently. For example, small planets at low signal to noise will typically remain as candidates rather than being confirmed or validated, and certain difficult FP scenarios such as transiting brown dwarfs are unlikely to be routinely recognised. In each case, such objects are likely to be more heavily represented in the unknown, non-dispositioned set.

For planets, we have good coverage of the planet set as a whole, as this is the entire confirmed planet training set, but planets in regions of parameter space where \Kepler has poor sensitivity should be viewed with suspicion.

For FPs, our training set includes a large number of non-astrophysical TCEs, giving good coverage of that scenario. For astrophysical FPs, we first make sure our training set is as representative as possible by including externally flagged FPs from \citet{Santerne:2016fz}. These cases use additional spectroscopic observations to mark candidates as FPs. However the bulk of small \Kepler candidates are not amenable to spectroscopic followup due to their stellar brightness. Utilising the FP flags in the archive as an indicator of the FP scenario, we have 2147 FPs showing evidence of stellar eclipses, and 1779 showing evidence of centroid offset and hence background eclipsing sources. 1087 show ephemeris matches, an indicator of a visible secondary eclipse and hence a stellar source. As such we have a wide coverage of key FP scenarios. 

It would be ideal to probe scenario by scenario and test the models in this fashion. Future work using specific simulated datasets will be able to explore this in more detail. Rare and difficult scenarios such as background transiting planets and transiting brown dwarfs are likely to be poorly distinguished by our or indeed any comparable method. In rare cases such as background transiting planets, which typically have transits too shallow to be detected, the effect on our overall results will be minimal. We note that these issues are equally present for currently utilised validation methods, and \texttt{vespa} for example cannot distinguish transiting brown dwarfs from planets \citep{Morton:2016ka}.

\section{Results}
\label{sectResults}
Our classification results are given in Table \ref{tabTCEresults}. The Table contains the classifier outputs for each TCE, calibrated if appropriate, as well as the relevant priors and final posterior probabilities adjusted by the priors. Several warning flags are included representing outliers, evolved host stars and detected close companions. Table \ref{tabKOIresults} shows the subset of Table \ref{tabTCEresults} for KOIs, and includes KOI specific information and \texttt{vespa} probabilitiies calculated for DR25.

\begin{table*}[h]
\caption{Full data table available online. This table describes the available columns.}
\label{tabTCEresults}
\begin{tabular}{lr}
\hline
\hline
Column & Description \\
\hline
tce\_id & Identifier composed by (KIC ID)\_(TCE planet number) \\
GPC\_score & Score from the GPC before priors are applied \\
MLP\_score & Calibrated score from the MLP model before priors are applied \\
RFC\_score & Calibrated score from the RFC before priors are applied \\
ET\_score & Calibrated score from the ET model before priors are applied \\
PP\_GPC & Planet probability from the GPC including priors \\
PP\_RFC & Planet probability from the RFC including priors \\
PP\_MLP & Planet probability from the MLP model including priors \\
PP\_ET & Planet probability from the ET model including priors \\
planet & Normalised prior probability for the planet scenario \\
targetEB & Normalised prior probability for the eclipsing binary on target scenario \\
targetHEB & Normalised prior probability for the hierarchical eclipsing binary scenario \\
targetHTP & Normalised prior probability for the hierarchical transiting planet scenario \\
backgroundBEB & Normalised prior probability for the background eclipsing binary scenario \\
backgroundBTP & Normalised prior probability for the background transiting planet scenario \\
secondsource & Normalised prior probability for any FP scenario on a known other stellar source \\
nonastro & Normalised prior probability for the non-astrophysical/systematic scenario \\
Binary & \citet{Berger:2018il} binarity flag (0=no evidence of binarity) \\
State & \citet{Berger:2018il} evolutionary state flag (0=main sequence, 1=subgiant, 2=red giant) \\
gaia & Flag for new Gaia DR2 sources within 25\arcsec bright enough to cause the signal (Section \ref{sectPosProbs}) \\
roboAO & Flag for robo-AO detected sources from \citet{Ziegler:2018ei} bright enough to cause the signal \\
MES & Multiple event statistic for the TCE. Results are valid for $MES>10.5$ \\
outlier\_score\_LOF & Outlier score using local outlier factor on whole dataset (Section \ref{sectOutliers}) \\
outlier\_score\_IF & Outlier score using isolation forest focused on training set (Section \ref{sectOutliers}) \\
class & Training set class, if any. $0=$ confirmed planets, $1=$ astrophysical FPs, $2=$ non-astrophysical FPs \\
\hline
\end{tabular}
\end{table*}

\subsection{Previously dispositioned objects}

To sanity check our method we consider the results of already dispositioned TCEs. For this testing we focus on the GPC results. There are two planets in the confirmed training set which score $<0.01$ in the GPC after applying the prior information. These are KOI2708.01 and KOI00697.01. Despite being labelled as confirmed in the NASA Exoplanet Archive KOI2708.01 is actually a certified FP, due a high level period match. This status is reflected in the positional probabilities, which give a relative probability of zero that the TCE originates from the host star. KOI00697.01 also has a positional probability indicating that the transit actually arises from a background star with high confidence, $>0.9999$. It is clear that both KOIs should be labelled FP.

There is also one KOI labelled as a FP which gains a score of $>0.99$ in the GPC, KOI3226.01. This KOI has a flag raised for having a `not-transit-like' signal. Visual inspection of the KOI shows stellar variability on a similar level to the transit signal, which may be distorting the transit signal on a quarter-by-quarter basis. The transits are however still evident in the lightcurve, and do not otherwise appear suspicious. We do not validate KOI03226.01, but our results indicate that its disposition may need to be reconsidered.

\subsection{Non-KOI TCEs}
\label{sectResnonKOI}
We additionally consider high scoring TCEs which are not in the KOI list to see if any merit further consideration. Nine TCEs score $>0.99$ in the GPC while passing our other checks. In each case the TCE was associated with the secondary eclipse of another TCE. For these TCEs, the \Kepler transiting planet search found the first TCE which was removed and the lightcurve searched again. In these cases the secondary eclipse of the original TCE remained in the lightcurve, and was `discovered' as an additional TCE. It appears metrics such as secondary eclipse depth were calculated after removing the primary eclipse, and so these `secondary' TCEs give all the indications of being planetary candidates. Such TCEs do not become KOIs and so would not be in danger of being mislabelled as validated planets. They highlight the dangers of poor information, in this case erroneous secondary eclipse measurements, both to our and other validation methods.

Overall the non-KOI TCEs have a mean GPC derived planet probability of 0.018, and a median of 0.002, as expected given these were not considered viable KOIs.

\subsection{Dependence on candidate parameters}
We investigate our model dependence on candidate parameters using the KOI list, discounting outliers as described in Section \ref{sectOutliers} but including KOIs with other warning flags. We focus on the planet probability as calculated by the GPC.

\begin{figure}
\resizebox{\hsize}{!}{\includegraphics{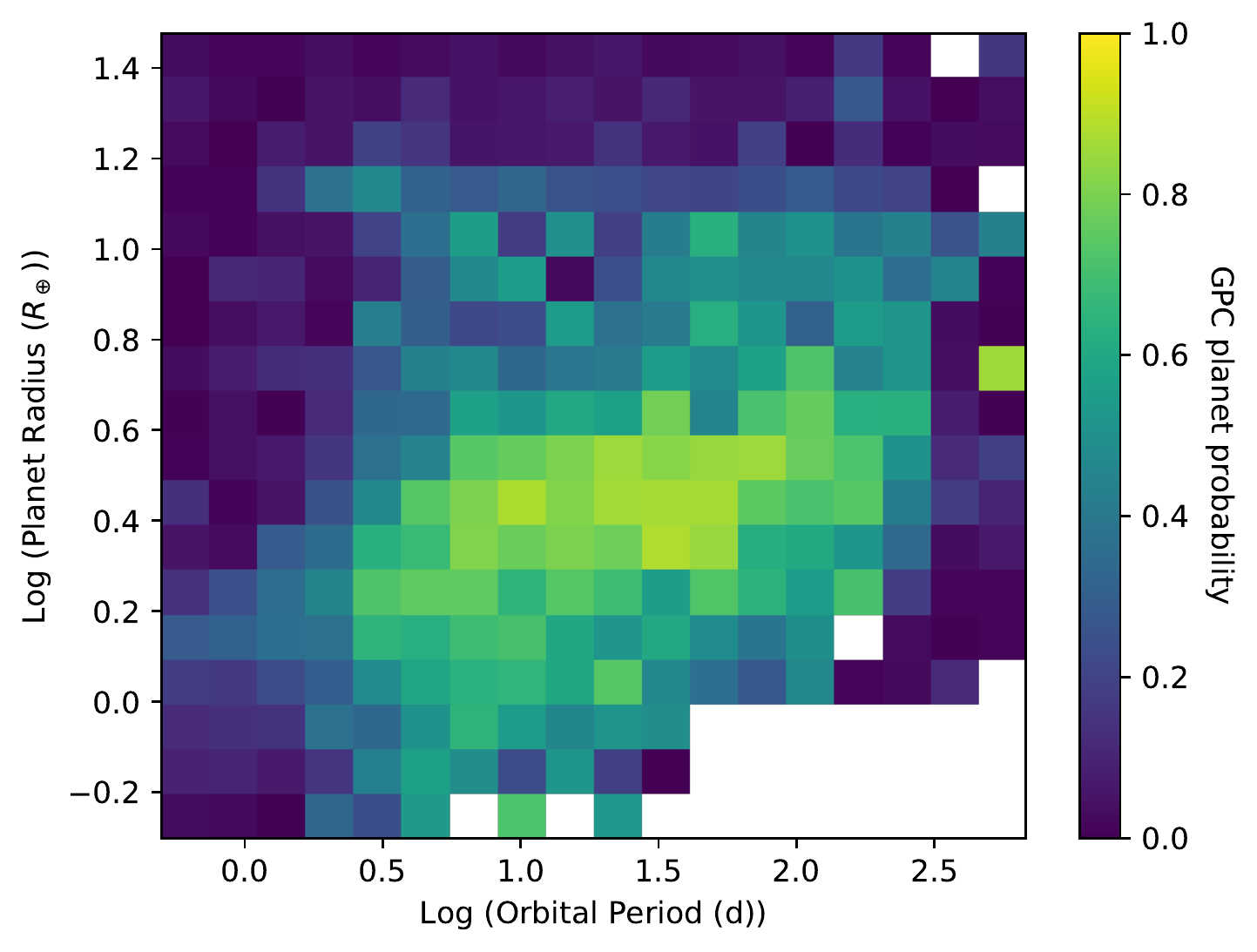}}
\caption{Mean GPC planet probability for KOIs binned in log planet radius and orbital period. Bins with no TCEs are white. Giant planets, and those at particularly long or short periods, are more likely to be classed as FPs. The GPC has more confidence in candidates in well-populated regions of parameter space, and loses confidence on average in KOIs which are near the limits of the \Kepler sensitivity in the lower right section of the figure.}
\label{figRPsensitivity}
\end{figure}

Table \ref{tabparamcomparison} shows the average planet probability including priors for KOIs based on planet radius and multiplicity, and demonstrates that KOIs in multiple systems score highly as would be expected from past studies of the effect of multiplicity on \Kepler FP occurrence rates. The high score of KOIs in multiple systems occurs despite no information on multiplicity being passed to the models. Table \ref{tabparamcomparison} also shows that the GPC planet probability decreases for giant planets, in agreement with previous studies showing the rate of FPs is larger for giant planet candidates \citep{Santerne:2016fz}. Figure \ref{figRPsensitivity} shows the median scores for KOIs of different radii and orbital period.

\begin{table}
\caption{GPC scores by KOI radius and multiplicity}
\label{tabparamcomparison}
\begin{tabular}{lllr}
\hline
\hline
Selection & Number	& \multicolumn{2}{c}{GPC P\_Planet}\\
&& Mean & Median \\
\hline
All & 7048 & 0.474 & 0.379 \\
KOI singles & 6148 & 0.296 & 0.013\\
KOIs in multiple systems & 1906 & 0.825 & 0.994 \\
${R}_{p}>= 15\;{R}_{\oplus }$ & 1558 & 0.029 & 0.006 \\
$10\;{R}_{\oplus }<= {R}_{p}< 15\;{R}_{\oplus }$	 & 323 & 0.295 & 0.063\\
$4\;{R}_{\oplus }<= {R}_{p}< 10\;{R}_{\oplus }$ & 824 & 0.351 & 0.029\\
$2\;{R}_{\oplus }<= {R}_{p}< 4\;{R}_{\oplus }$ & 2482 & 0.666 & 0.982\\
${R}_{p}< 2\;{R}_{\oplus }$ & 2867 & 0.456 & 0.360\\
\hline
\end{tabular}
\end{table}

\subsection{Comparison to \citet{Santerne:2016fz}}
\citet{Santerne:2016fz} provided dispositions of some \Kepler candidates using independent data. The mean and median planet probabilities from the GPC are shown for each disposition type in \citet{Santerne:2016fz} in Table \ref{tabsantcomparison}, including planets, brown dwarfs (BD), eclipsing binaries (EB), contaminating eclipsing binaries (CEB) and unknowns.

We achieve a high score for planets and low scores for EBs and CEBs. BDs are also scored highly, indicating we are insensitive to that FP scenario similarly to \texttt{vespa} \citep{Morton:2016ka}, although in our case the BDs score lower and well below the validation threshold. Typically our model is less confident of giant planets (Table \ref{tabparamcomparison}) and this guards against the inaccurate validation of brown dwarfs. We hypothesis that this is also why the \citet{Santerne:2016fz} planets score relatively lower than the general confirmed planet case, as they are larger than the average KOI.

\begin{table}
\caption{GPC planet probabilities for \citet{Santerne:2016fz} dispositioned KOIs}
\label{tabsantcomparison}
\begin{tabular}{llrr}
\hline
\hline
Selection	& Number	& \multicolumn{2}{c}{GPC GPC P\_Planet}\\
	&&Mean & Median \\
\hline
Planets & 44 & 0.711& 0.808\\
EB & 48 & 0.166 & 0.100\\
CEB & 15 & 0.163 & 0.045\\
BD & 3 & 0.908 & 0.910 \\
Unknown & 18 & 0.688 & 0.717 \\
\hline
\end{tabular}
\end{table}

\subsection{Comparison to \texttt{vespa}}
\label{sectVespaComp}
\begin{figure}
\resizebox{\hsize}{!}{\includegraphics{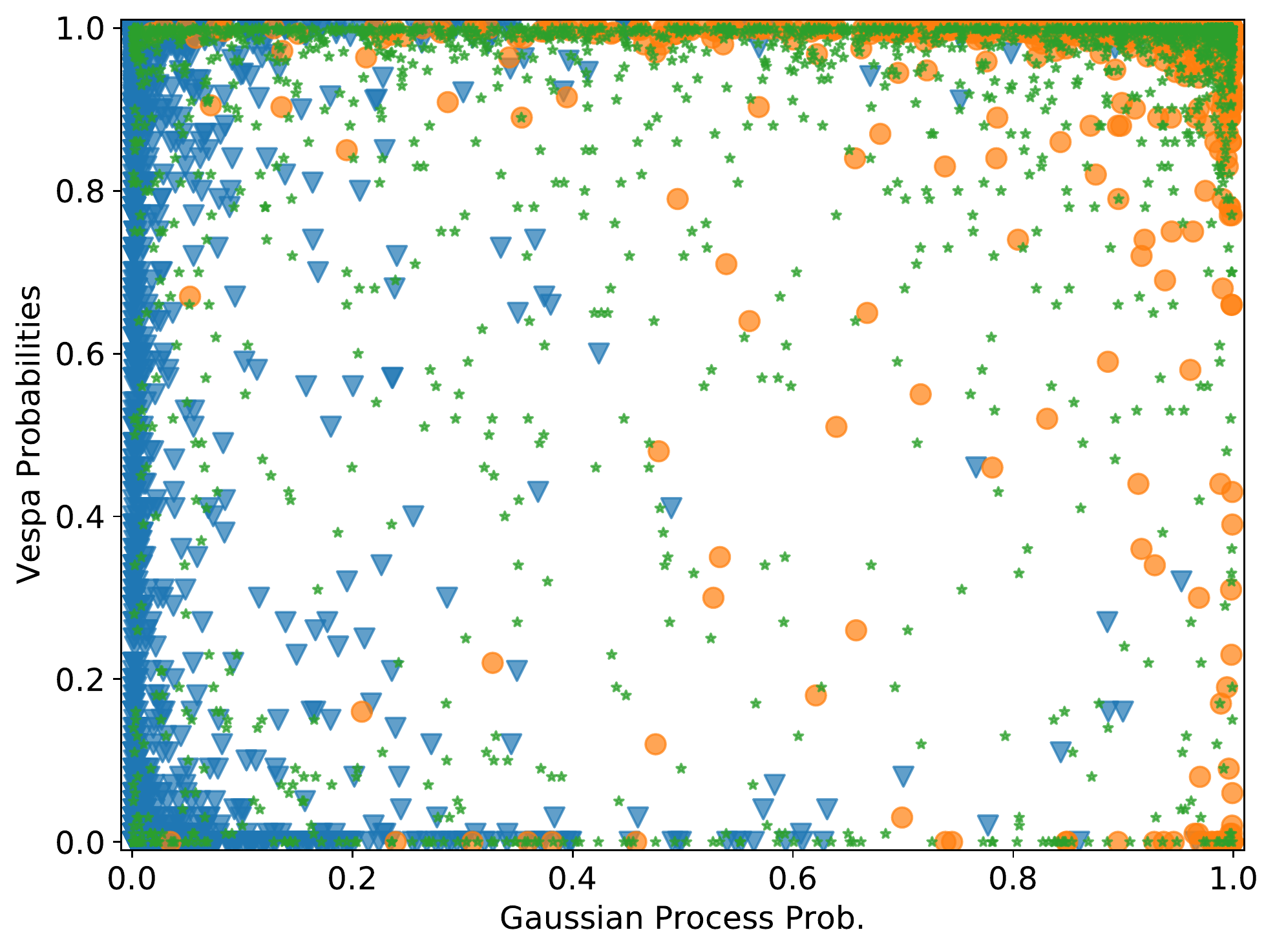}}
\caption{Comparison of GPC scores before application of priors and \texttt{vespa} false positive probabilities. We plot $1-FPP_\textrm{vespa}$ to allow direct comparison. Confirmed planets are orange circles, false positives are blue traingles, and undispositioned candidates are green stars. Significant divergence is seen, although the GPC and \texttt{vespa} agree in 73\% of cases.}
\label{figvespacomp}
\end{figure}

Figure \ref{figvespacomp} shows the GPC scores as compared to the FPP calculated by \texttt{vespa} \citep{Morton:2016ka}. We use the updated vespa false positive probabilities (FPPs) available at the NASA Exoplanet Archive for DR25, and consider only KOIs which pass our outlier checks. The DR25 \texttt{vespa} FPP scores have not been published in their own paper and hence have not been used to update planet dispositions in the Exoplanet Archive, despite being available there. GPC scores are plotted before application of prior information to allow a more direct comparison, as the \texttt{vespa} probabilities available on the NASA exoplanet archive appear not to include updated positional probability information. Although the plot appears remarkably divergent we highlight that in 73\% percent of cases the classification is the same using a threshold of 50\%. Both methods tend to confidently classify candidates as planets or FPs, with intermediate values sparsely populated. Furthermore, candidates which do receive intermediate scores show no correlation between the methods. As such we caution against using such intermediate candidates for occurrence rate studies, even if weighting by the GPC score or \texttt{vespa} FPP would appear to be statistically valid.

\begin{figure}
\resizebox{\hsize}{!}{\includegraphics{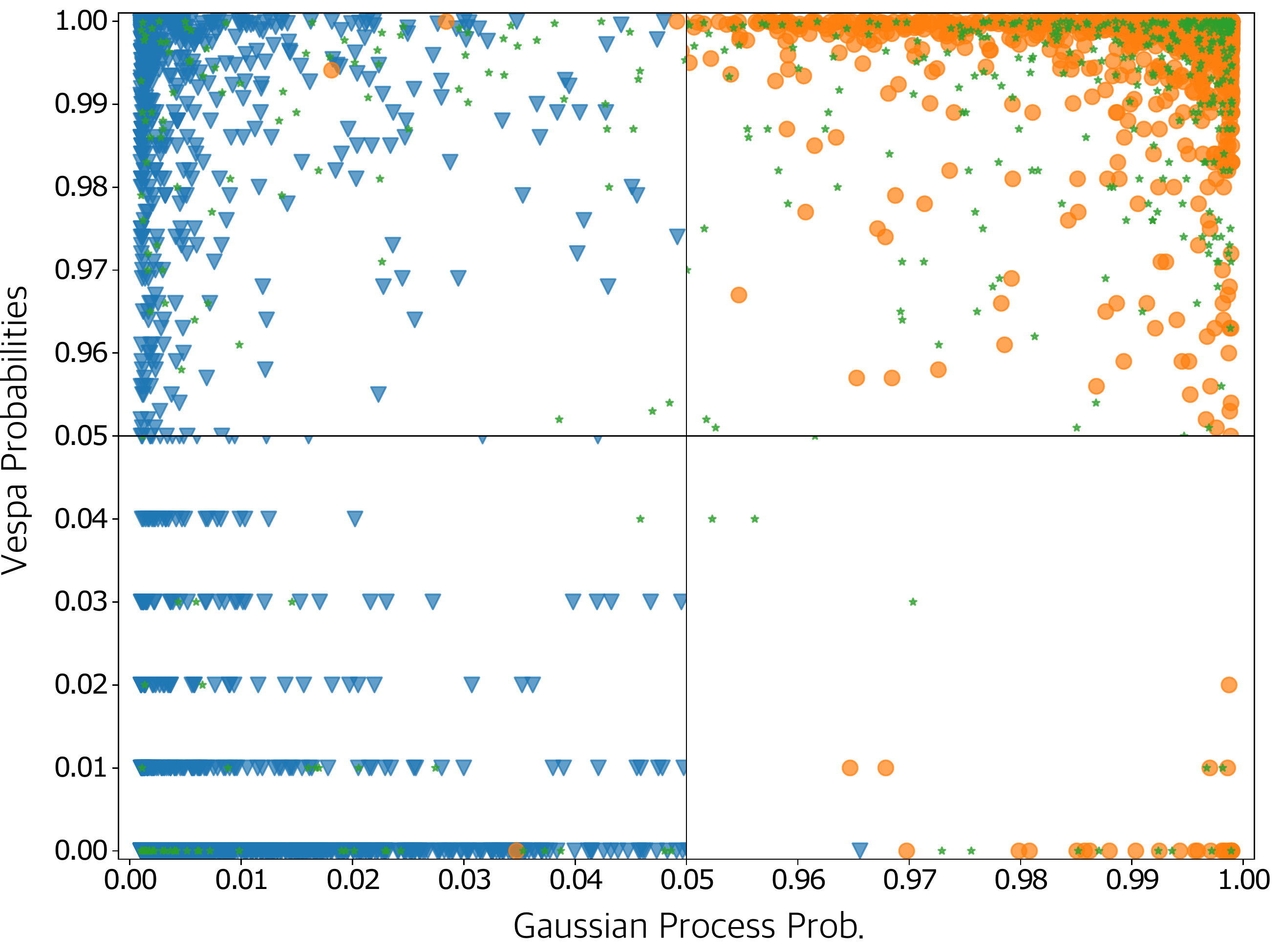}}
\caption{As Figure \ref{figvespacomp} showing a zoom of each corner. The banding in the lower left panel is due to the reported precision of \texttt{vespa} results.}
\label{figvespacomp_zoom}
\end{figure}

The methods also strongly disagree in a small but significant number of cases. Figure \ref{figvespacomp_zoom} shows a zoom of each corner of Figure \ref{figvespacomp}. The GPC gives 31 non-outlier KOIs a probability $>=0.99$ of being a planet where the \texttt{vespa} FPP shows a false positive probability of $>=0.99$, 24 of which are confirmed planets. In the other corner, the GPC classifies 399 non-outlier KOIs as strong FPs (probability $<=0.01$) where the \texttt{vespa} FPP shows a false positive probability $<=0.01$, apparently validating them. 375 of these KOIs are designated FPs. For these cases our GPC appears to be more reliable, potentially as it is trained on the full \Kepler set of FPs rather than limited to specific scenarios which may not fully explore unusual cases, or reliably account for the candidate distributions in the \Kepler candidate list. A study of some of these discrepant cases in detail did not reveal any typical mode for these \texttt{vespa} failures, and included clear stellar eclipses, centroid offsets, ghost halo pixel-level systematics, and ephemeris matches. Overall the comparison highlights the value of independent methods for planet validation, and we recommend extreme caution is used when validating planets, ideally avoiding using a single method.

\subsection{Inter-model comparison}
\label{sectmodelcomp}
\begin{figure}
\resizebox{\hsize}{!}{\includegraphics{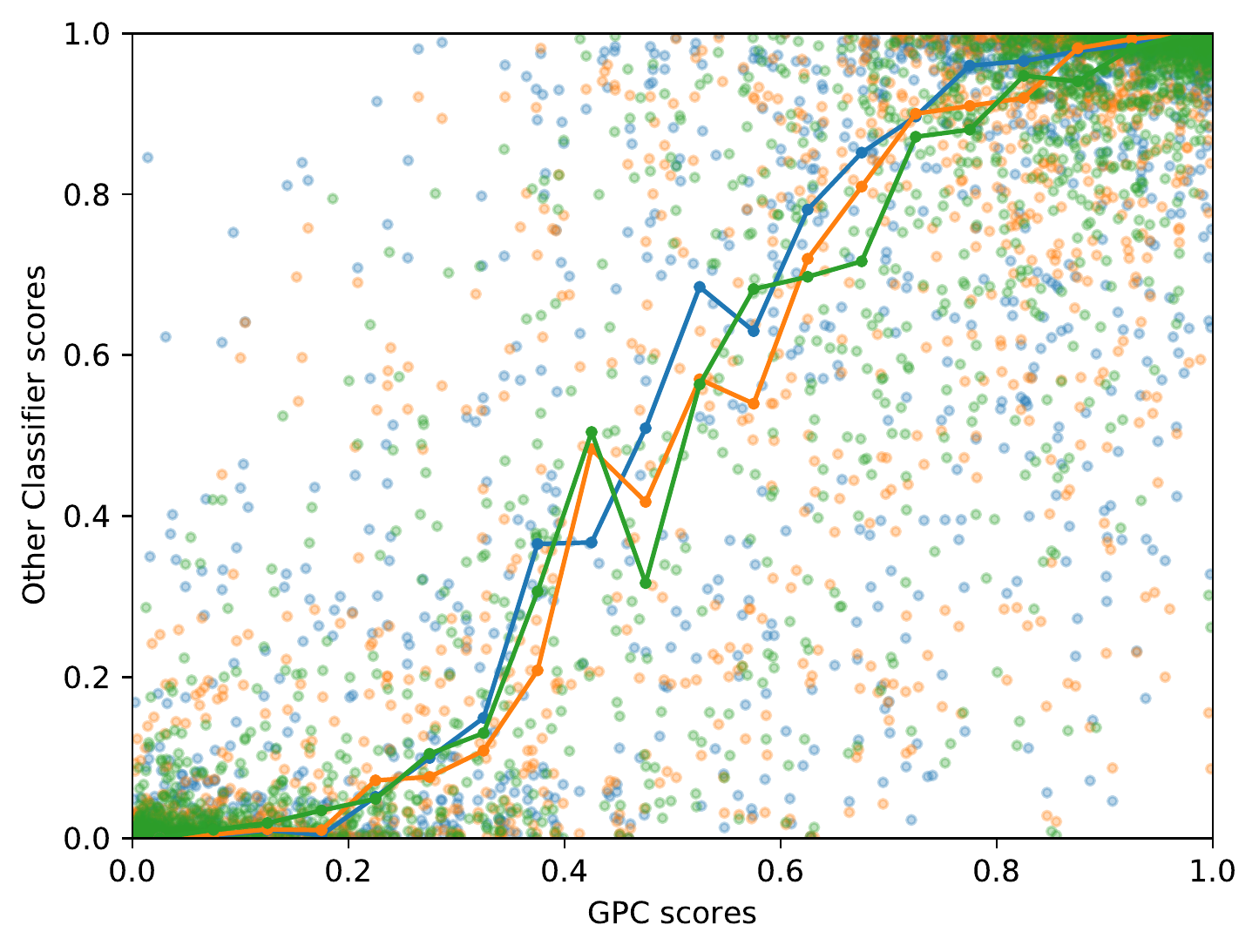}}
\caption{GPC scores as compared to the RFC (blue), ET (orange) and MLP (green). The median scores of 20 evenly distributed bins are overplotted. The GPC is typically more conservative when making classifications than the other models, leading to the visible trend.}
\label{figmodelcomp}
\end{figure}

As our framework considers four separate models we can compare the results of these. Figure \ref{figmodelcomp} shows the output classifier scores before applying prior probabilities for KOIs which pass our outlier checks. Although the models typically agree on a classification, there is still significant spread in the exact values, and the GPC in particular tends to be more conservative in its classifications than the other classifiers, as it is an inherently probabilistic framework and so more comprehensively considers probabilities across the range. Spread in intermediate values is expected, as the probability calibration is known to be poorly determined there due to a small number of samples. The observed spread highlights the importance of only validating planets where all models agree, and the dangers in building machine learning planet validation tools relying on only one classifier.

\subsection{Newly validated planets}
We set stringent criteria to validate additional TCEs as planets. We require each of the four classifiers to cross the standard validation threshold of 0.99, representing a less than 1\% chance of being a FP. The TCE must also be a KOI, the evolutionary state flag from \citet{Berger:2018il} must not show a subgiant or giant host star, both outlier flags must not be set, the binary flag from \citet{Berger:2018il} derived from Gaia DR2 and robo-AO must not show evidence for a binary, there must be no sufficiently bright new detected Gaia DR2 or robo-AO companions as described in Section \ref{sectPosProbs}, the MES of the TCE must be greater than 10.5 to avoid a high systematic false alarm chance, and the score representing the quality of the positional probabilities calculated in \citet{Bryson:tz} must be higher than 0.3 indicating a good positional fit. Although the positional probability that the host star is the source of the transit signal is incorporated in our priors, we additionally require validated KOIs to have a relative probability of being on the target star of at least 0.95.

\begin{figure}
\resizebox{\hsize}{!}{\includegraphics{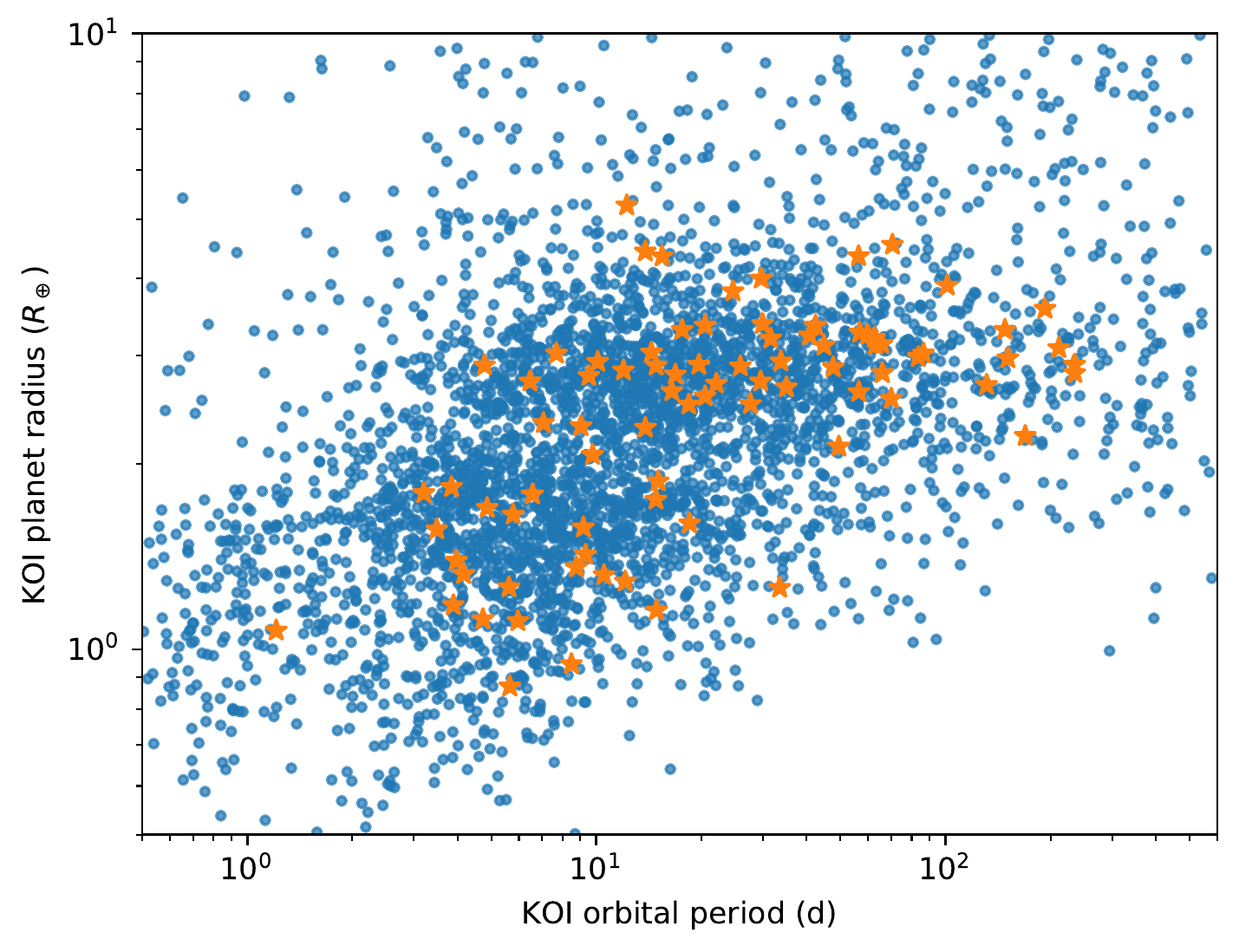}}
\caption{KOIs which pass our validation steps (orange stars) in the context of \Kepler candidate and confirmed planets (blue dots). Known FPs are not plotted.}
\label{fignewvalRP}
\end{figure}

83 KOIs passed these criteria and were not already validated or confirmed. As a sanity check we reran \texttt{vespa} on these objects using our transit photometry and GAIA parallax information. 50 of the 83 KOIs obtained a less than 1\% chance of being a FP using these updated \texttt{vespa} results. Given the above discrepancies between our models and \texttt{vespa}, we take the cautious approach of only newly validating planets which agree between both methods, at least until the discrepancies are more fully understood. We do note that in Section \ref{sectVespaComp}, serious discrepancies between our models and \texttt{vespa} were almost entirely resolved in favour of our models by the independent \Kepler pipeline designations. The 50 validated planets are listed in Table \ref{tabnewvals}, and the 33 candidates for which there is still disagreement in Table \ref{tabnewnotvals}. 15 of the 83 high-probability candidates are in systems which already host a confirmed or validated planet. We plot the 83 planets in Figure \ref{fignewvalRP} against the context of known \Kepler planets and candidates.

\begin{table*}
\caption{New validated planets. Full table available online.}
\label{tabnewvals}
\begin{threeparttable}
\begin{tabular}{lllllllllllr}
\hline
\hline
Planet & KOI & KIC & Period & $R_p$ & GPC & RFC & MLP & ET & vespa\_fpp\tnote{1} & P target\tnote{2} & Pos. score\tnote{2} \\ 
 &  &  $d$  &  $R_\oplus$  &  &  &  & &  & & & \\
\hline
Kepler-1663b & K00252.01 & 11187837 & 17.605 & 3.30 & 0.9985 & 1.0000 & 0.9994 & 1.0000 & 0.00001 & 1.0 & 1.0 \\
Kepler-1664b & K00349.01 & 11394027 & 14.387 & 3.03 & 0.9985 & 1.0000 & 0.9971 & 1.0000 & 0.00492 & 1.0 & 1.0 \\
Kepler-598c & K00555.02 & 5709725 & 86.494 & 3.02 & 0.9994 & 0.9986 & 1.0000 & 1.0000 & 0.00989 & 1.0 & 1.0 \\
Kepler-1665b & K00650.01 & 5786676 & 11.955 & 2.84 & 0.9985 & 1.0000 & 0.9990 & 1.0000 & 0.00288 & 1.0 & 1.0 \\
Kepler-647c & K00691.01 & 8480285 & 29.666 & 4.00 & 0.9995 & 1.0000 & 1.0000 & 1.0000 & 0.00000 & 1.0 & 1.0 \\
Kepler-716c & K00892.02 & 7678434 & 3.970 & 1.39 & 0.9904 & 0.9916 & 0.9993 & 0.9980 & 0.00574 & 1.0 & 0.44 \\
\vdots &  \vdots& \vdots  & \vdots  & \vdots & \vdots &  \vdots& \vdots& \vdots & \vdots& \vdots&\vdots \\
\hline
\end{tabular}
\begin{tablenotes}
\item [1] New \texttt{vespa} FPP values calculated using the Gaia parallax and our photometry. Validated planets should have a low value, unlike the other columns.
\item [2] Probability on target and positional score from \citet{Bryson:tz}
\end{tablenotes}
\end{threeparttable}
\end{table*}

\begin{table*}
\caption{KOIs with $>0.99$ probability of being a planet from our models where an updated vespa calculation does not agree. Full table available online.}
\label{tabnewnotvals}
\begin{threeparttable}
\begin{tabular}{llllllllllr}
\hline
\hline
KOI & KIC & Period & $R_p$ & GPC & RFC & MLP & ET & vespa\_fpp\tnote{1} & P target\tnote{2} & Pos. score\tnote{2} \\ 
 &  &  $d$  &  $R_\oplus$  &  &  &  & &  &\\
\hline
K00092.01 & 7941200 & 65.705 & 3.13 & 0.9933 & 0.9991 & 0.9979 & 1.0000 & 0.19300 & 1.0 & 1.0\\
K00247.01 & 11852982 & 13.815 & 2.29 & 0.9988 & 1.0000 & 0.9984 & 1.0000 & 0.02470 & 1.0 & 1.0\\
K00427.01 & 10189546 & 24.615 & 3.81 & 0.9917 & 1.0000 & 0.9968 & 1.0000 & 0.10000 & 1.0 & 0.5\\
K00599.01 & 10676824 & 6.454 & 2.72 & 0.9985 & 1.0000 & 0.9981 & 1.0000 & 0.05210 & 1.0 & 1.0\\
K00704.01 & 9266431 & 18.396 & 2.50 & 0.9986 & 0.9997 & 0.9984 & 0.9998 & 0.01210 & 1.0 & 1.0\\
K00810.01 & 3940418 & 4.783 & 2.89 & 0.9994 & 0.9966 & 0.9999 & 1.0000 & 0.03300 & 1.0 & 1.0\\
\vdots &  \vdots& \vdots  & \vdots  & \vdots & \vdots &  \vdots& \vdots& \vdots & \vdots& \vdots\\
\hline
\end{tabular}
\begin{tablenotes}
\item [1] New \texttt{vespa} FPP values calculated using the Gaia parallax and our photometry. Validated planets should have a low value, unlike the other columns.
\item [2] Probability on target and positional score from \citet{Bryson:tz}
\end{tablenotes}
\end{threeparttable}
\end{table*}

\section{Discussion}
\label{sectDiscuss}
Statistical exoplanet validation remains a key part of the exoplanet discovery process, originally to accommodate faint \Kepler host stars where the potential for independent follow-up was low but continuing with \TESS to aid discovery for difficult or time consuming candidates \citep{Quinn:2019cr,Vanderburg:2019di}. The planet validation literature is dominated by a small number of methods, only one of which, \texttt{vespa}, is regularly used, relatively fast to run and available publicly. An accurate set of confirmed and validated exoplanets is crucial for many fields of research, including planet architecture, formation and population synthesis. As such, alternative validation methods which are fast to run are highly valuable. Our methodology has been demonstrated on the \Kepler dataset and will be developed to work more generally, particularly with \TESS data.

\subsection{Caveats and Limitations}

There are several implicit assumptions and limitations to our methodology, which we summarise here.

\begin{itemize}

\item \textbf{Training set coverage}
A fundamental part of the machine learning models we are using is the use of a training set. By using this set, we are implicitly assuming that candidates are not overwhelmingly members of a scenario not represented in the FP training set, or some other rare case that is not expected or understood. The coverage of the training set for typical scenarios is discussed in Section \ref{sectTsetCoverage}.

\item \textbf{Quality of input data and parameter accuracy} 
All inputs to the models are treated in the same way, and hence systematic biases in the input data are not critically important where they affect all candidates equally. Biases associated with one particular scenario are similarly not critical if they affect all instances of that scenario equally. Issues with single candidates are however a potential problem, and an individual tested candidate with bad input values may lead to bad scores.

\item \textbf{Reliance on previous dispositions}
Our method builds on previous efforts to disposition \Kepler candidates as planets or FPs. In particular a large fraction of the known \Kepler planets come from \texttt{vespa} validations, which means we are potentially building in the same biases. This effect is mitigated by also using non-\texttt{vespa} dispositions and updating our inputs with the latest Gaia, \Kepler and positional probability results, and Section \ref{sectVespaComp} shows we are producing independent results to \texttt{vespa}. Further, we are implicitly assuming that the majority of confirmed planets really are planets, and the same for FPs.

\item \textbf{Outlier or anomalous candidates} Our models are only valid for candidates which lie in well represented parts of the input parameter space. Scores for outliers are potentially invalid. This is a clear but well understood limitation, and outliers are flagged in Section \ref{sectOutliers}.

\item \textbf{Calibration precision}
Three of our four models require a probability calibration step, which is only as precise as the number of samples used for calibration allows. We have reached theoretical precisions of 1\% in the 0--0.01 and 0.99--1 probability ranges, but intermediate values should be treated with caution, as most tested candidates are given scores at the extreme ends of the scale. Any reader looking to use intermediate scores for their work should use the GPC results only which do not depend on calibration, and should take care in any case given the discussion in Section \ref{sectVespaComp}.

\item \textbf{Planet multiplicity}
No adjustment has been made for candidates in multiple systems, due to the complexity of the resulting probabilities and the difficulty of ascertaining how many TCEs on a given star are likely planets, and hence should be counted in any multiplicity effects. However, the models do find higher scores for candidates in multiple systems, even without applying a `multiplicity boost'.

\item \textbf{Specificity to \Kepler}
This work is built and tested for \textit{Kepler}, and we warn against casual application to other datasets such as \textit{TESS}. The method should work  in principle but detailed care needs to be taken to work with the above limitations, and build a suitable training set. In this work we can use the actual distribution of \Kepler discoveries; for future less mature missions care must be taken when simulating training sets to use appropriate distributions. We also use outputs of the \Kepler pipeline, which would be hard to exactly recreate. Nonetheless, it should be relatively simple to create statistics containing the same information, such as tests of the secondary eclipse depth, for other missions, and such statistics are standard outputs of most current vetting procedures \citep{Kostov:2019ij}. 

\item \textbf{Inter-model divergence}
Section \ref{sectmodelcomp} showed that our four models show a significant spread in output probabilities for a given KOI. It is important in future similar work to use multiple models to confirm a validation decision and guard against over-reliance on a single model. The divergence also highlights that intermediate FPP values should be treated with caution, as already evident by the comparison with \texttt{vespa} and the calibration issues discussed above.

\end{itemize}


\subsection{Comparison to other methods}

There are two lines of past work relevant to our method. The first is previous efforts at planet validation, the key comparable example of which is the \texttt{vespa} algorithm. Our results were compared to \texttt{vespa} in Section \ref{sectVespaComp}, but we discuss the methodological differences here. In particular, \texttt{vespa} uses a least squares fit of a trapezoid model to the TCE lightcurve to perform scenario model comparison between several defined planet and FP scenarios, in combination with stellar parameters and other auxiliary information. In our method the model comparison is performed by the machine learning algorithms, with the models defined by the input training set. Our lightcurve representation is more complex, being either a SOM-based dimensional reduction of the lightcurve or a direct binned view of the transit, depending on the model used. We use the same auxiliary data, bolstered by other outputs of the \Kepler transiting planet search as detailed in Section \ref{sectData}. Particular additions over \texttt{vespa} include pixel level diagnostics such as ghost halo issues, detailed information on transit shape capable of identifying known systematic shapes, and ephemeris matches.

Our method incorporates several improvements available due to recently released datasets or new understanding of the key issues. In particular we incorporate the non-astrophysical false positive scenario directly in the model comparison by including a large group in our training set, accounting for systematic false alarms as warned in \citet{Burke:2019js}. \texttt{vespa} as described in \citet{Morton:2016ka} accounted for non-astrophysical false alarms using a statistic calculated on the transit shape which was applied separately. Our models also run extremely quickly, and can classify the entire TCE catalogue of \mytilde 34000 candidates in minutes on a typical desktop once trained and auxiliary data calculated. We have included the latest Gaia DR2 information on the host stars and blended companions, and the latest catalogue of robo-AO detected companions \citep{Ziegler:2018ei}.

The other less common but still actively used planet validation algorithm is \texttt{PASTIS} \citep{Santerne:2015bb,Diaz:2014kd}. \texttt{PASTIS} performs model comparison via direct MCMC fits to potentially multi-color lightcurve data and the stellar spectral energy distribution considering each false positive scenario in turn, and as such is the gold standard. The downside is that \texttt{PASTIS} is slow to run and can only be applied to individual candidates in some cases. Our model is much faster to run, although simplified.

The second line of comparison is previous attempts to classify planet candidates as FPs using machine learning methods. With the advent of large datasets such work is increasingly common, and classifiers have been built for \Kepler \citep{McCauliff:2015fb,Shallue:2018jy,Ansdell:2018dq,Caceres:2019ju}, K2 \citep{Armstrong:2017cp,Dattilo:2019ht}, \TESS \citep{Yu:2019ba,Osborn:2019wo}, NGTS \citep{Armstrong:2018ey,Chaushev:2019gx}, and WASP \citep{Schanche:2018kl}. For the \Kepler dataset, \citet{Caceres:2019ju} built a random forest model to find good candidates among the results from their `autoregressive planet search' algorithm, achieving an AUC of 0.997 in classifying planet candidates against false positives. \citet{Shallue:2018jy} used a convolutional neural net for a similar purpose, achieving an AUC of 0.988 and again aiming to separate candidates from false positives using a different planet search method. Measured by AUC, the best past performance on \Kepler candidates was in \citet{McCauliff:2015fb}, who achieved an AUC of 0.998 using an RFC when separating planets from FPs of any type. The key step we take in this work beyond those or other previous attempts to identify planets among candidate signals is to focus on separating true planets, as opposed to just planetary candidates, from FPs in the candidate set probabilistically. We also introduce a GPC for exoplanet candidate vetting for the first time. Although our goals are different and so not strictly comparable, the AUC metrics from our GPC, RFC, ET and MLP models are 0.999, 0.999, 0.999 and 0.998 respectively, when separating confirmed planets from false positives.

\subsection{Future Work}

For both planets and false positives, we hypothesise that a rigorous set of simulated objects will allow detailed model testing and improved training with increased training set size and coverage, and intend to introduce these improvements in a later work. Such a sample will allow detailed scenario by scenario comparison and give a deeper understanding of the strengths and weaknesses with respect to specific scenarios. Utilising the direct distribution of discovered \Kepler planets and FPs does however have the advantage that the distribution is available to inform our models, implying that difficult to distinguish FP scenarios that are nonetheless intrinsically rare will not bias the results.

In line with utilising simulated training sets, we intend to build a codebase to make the method publicly accessible. We have not made the code from this work public as it is specific to the \Kepler pipeline and DR25 data release, the results for which we publish here. We aim to release a more general code applicable to \TESS or other mission data in future.

\section{Conclusion}
We have developed a new planet validation framework utilising several machine learning models. Our method has proved successful and able to validate planets rapidly. The potential use cases extend beyond planet validation to candidate vetting and prioritisation, crucial given the data rate of current and upcoming surveys.

This work represents the first time to our knowledge that a large scale comparison of validation methods, specifically to the popular \texttt{vespa} algorithm, has been attempted. The resulting discrepancies seen in Section \ref{sectVespaComp} are concerning given the high fraction of known planets discovered using validation techniques. As a consequence, we strongly caution against validating planets in future with only one method, be it ours, \texttt{vespa}, or any other technique which is not a full Bayesian model of all the available information such as \texttt{PASTIS}. This caution should be taken extremely seriously when considering validation of multiple planets simultaneously, given the potential to distort the confirmed planet population if unrecognised biases exist.

\section*{Acknowledgements}
We would like to thank Chelsea Huang for conversations which helped this work. DJA acknowledges support from the STFC via an Ernest Rutherford Fellowship (ST/R00384X/1). This work was supported by The Alan Turing Institute for Data Science and AI under EPSRC grant EP/N510129/1. We gratefully acknowledge the support of NVIDIA Corporation with the donation of the Titan Xp GPU used for this research. This research has made use of the NASA Exoplanet Archive, which is operated by the California Institute of Technology, under contract with the National Aeronautics and Space Administration under the Exoplanet Exploration Program. This paper includes data collected by the Kepler mission and obtained from the MAST data archive at the Space Telescope Science Institute (STScI). Funding for the Kepler mission is provided by the NASA Science Mission Directorate. STScI is operated by the Association of Universities for Research in Astronomy, Inc., under NASA contract NAS 5–26555.

\section*{Data Availability}
All data used in this manuscript are available from the NASA Exoplanet Archive at https://exoplanetarchive.ipac.caltech.edu/. Newly generated data is available within the manuscript.

\bibliography{papers061219}
\bibliographystyle{mn2e_fix}


\section{Appendix}
\setcounter{table}{0}
\renewcommand{\thetable}{A\arabic{table}}
\begin{landscape}
\begin{table}
\caption{Full table available online.}
\label{tabKOIresults}
\begin{threeparttable}
\begin{tabular}{llllllllllllllllr}
\hline
\hline
KOI & Period & $R_p$\tnote{1} & GPC\tnote{2} & RFC\tnote{2} & MLP\tnote{2} &  ET\tnote{2} & \texttt{vespa}\_fpp\tnote{3} & \multicolumn{4}{c}{Flags} & Prob. on target\tnote{6} & posiscore\tnote{6} & MES & \multicolumn{2}{c}{Outlier score} \\
 & $d$ & $R_\oplus$ &  &  &  &  & & Binary\tnote{4} & State\tnote{4} & gaia & roboAO\tnote{5} & & & & LOF &IF\\
\hline
K00001.01 & 2.471 & 14.40 & 0.395 & 0.141 & 0.728 & 0.139 & 0.010 & 2 & 0 & 0 & 1 & 1.00 & 1.00 & 6468.0 & -1.12 & -0.55\\
K00002.01 & 2.205 & 17.11 & 0.071 & 0.402 & 0.433 & 0.198 & 0.000 & 0 & 0 & 0 & 0 & nan & 0.00 & 3862.0 & -1.29 & -0.55\\
K00003.01 & 4.888 & 4.99 & 0.272 & 0.824 & 0.993 & 0.762 & 0.000 & 0 & 0 & 0 & 0 & nan & 0.00 & 2035.0 & -1.64 & -0.54\\
K00004.01 & 3.849 & 14.01 & 0.239 & 0.330 & 0.125 & 0.044 & 0.028 & 2 & 1 & 0 & 0 & 1.00 & 1.00 & 235.6 & -1.17 & -0.48\\
K00005.01 & 4.780 & 8.94 & 0.366 & 0.094 & 0.303 & 0.031 & 0.160 & 0 & 1 & 0 & 0 & 0.29 & 1.00 & 360.2 & -1.06 & -0.50\\
K00006.01 & 1.334 & 1.09 & 0.000 & 0.000 & 0.000 & 0.000 & 0.000 & 0 & 0 & 0 & 0 & 0.00 & 1.00 & 21.5 & -1.01 & -0.49\\
K00007.01 & 3.214 & 4.54 & 0.995 & 0.997 & 1.000 & 0.998 & 0.000 & 0 & 1 & 0 & 0 & 1.00 & 1.00 & 294.5 & -1.08 & -0.50\\
K00008.01 & 1.160 & 1.12 & 0.000 & 0.000 & 0.000 & 0.000 & 0.000 & 0 & 0 & 1 & 0 & 0.00 & 1.00 & 36.6 & -1.08 & -0.52\\
K00009.01 & 3.720 & 6.24 & 0.000 & 0.000 & 0.000 & 0.000 & 0.990 & 0 & 0 & 0 & 0 & 0.00 & 1.00 & 579.9 & -1.13 & -0.50\\
K00010.01 & 3.522 & 16.11 & 0.706 & 0.856 & 0.647 & 0.969 & 0.001 & 0 & 1 & 0 & 0 & 1.00 & 1.00 & 1726.0 & -1.05 & -0.52\\
K00011.01 & 3.748 & 2.90 & 0.000 & 0.000 & 0.000 & 0.000 & 1.000 & 0 & 0 & 0 & 0 & 0.00 & 1.00 & 98.6 & -1.32 & -0.50\\
K00012.01 & 17.855 & 14.59 & 0.742 & 0.881 & 0.595 & 0.700 & 0.000 & 0 & 0 & 0 & 0 & 1.00 & 1.00 & 417.9 & -1.30 & -0.52\\
K00013.01 & 1.764 & 17.72 & 0.081 & 0.562 & 0.879 & 0.270 & 0.150 & 2 & 0 & 0 & 0 & nan & 0.00 & 6791.0 & -1.37 & -0.56\\
K00014.01 & 2.947 & 5.82 & 0.005 & 0.003 & 0.006 & 0.003 & nan & 0 & 0 & 1 & 0 & 0.01 & 0.12 & 318.8 & -1.35 & -0.49\\
K00015.01 & 3.012 & 9.20 & 0.000 & 0.000 & 0.000 & 0.000 & 1.000 & 0 & 0 & 0 & 0 & 0.00 & 1.00 & 372.4 & -1.46 & -0.52\\
K00016.01 & 0.895 & 4.74 & 0.000 & 0.000 & 0.000 & 0.000 & 0.000 & 0 & 0 & 0 & 0 & 0.00 & 1.00 & 202.3 & -1.42 & -0.55\\
K00017.01 & 3.235 & 12.87 & 0.883 & 0.981 & 0.991 & 0.997 & 0.001 & 0 & 0 & 0 & 0 & 1.00 & 1.00 & 2986.0 & -1.12 & -0.54\\
K00018.01 & 3.548 & 15.17 & 0.749 & 0.922 & 0.996 & 0.943 & 0.000 & 0 & 1 & 0 & 0 & 1.00 & 1.00 & 2269.0 & -1.13 & -0.54\\
K00019.01 & 1.203 & 10.85 & 0.370 & 0.003 & 0.019 & 0.005 & 0.990 & 0 & 1 & 0 & 0 & 1.00 & 1.00 & 1461.0 & -1.11 & -0.54\\
K00020.01 & 4.438 & 19.53 & 0.762 & 0.755 & 0.995 & 0.948 & 0.004 & 0 & 0 & 0 & 0 & 1.00 & 1.00 & 3303.0 & -1.11 & -0.54\\
K00021.01 & 4.289 & 18.32 & 0.000 & 0.000 & 0.000 & 0.000 & 1.000 & 0 & 0 & 0 & 0 & 0.00 & 1.00 & 563.3 & -1.12 & -0.51\\
K00022.01 & 7.891 & 13.23 & 0.917 & 0.994 & 0.973 & 0.999 & 0.005 & 0 & 0 & 0 & 0 & 1.00 & 1.00 & 1973.0 & -0.99 & -0.53\\
K00023.01 & 4.693 & 25.63 & 0.086 & 0.121 & 0.086 & 0.006 & 0.110 & 0 & 0 & 0 & 0 & 1.00 & 1.00 & 2740.0 & -1.02 & -0.54\\
K00024.01 & 2.086 & 9.39 & 0.000 & 0.000 & 0.000 & 0.000 & 0.095 & 0 & 0 & 0 & 0 & 0.00 & 1.00 & 698.0 & -1.38 & -0.53\\
K00025.01 & 3.133 & 45.81 & 0.052 & 0.001 & 0.001 & 0.000 & 0.940 & 0 & 1 & 0 & 0 & 1.00 & 1.00 & 1488.0 & -1.10 & -0.55\\
K00026.01 & 15.040 & 17.16 & 0.000 & 0.000 & 0.000 & 0.000 & 0.150 & 0 & 0 & 0 & 0 & 0.00 & 1.00 & 1196.0 & -1.34 & -0.55\\
K00027.01 & 1.142 & 78.72 & 0.014 & 0.000 & 0.000 & 0.000 & 0.045 & 0 & 1 & 0 & 0 & 1.00 & 1.00 & 6993.0 & -1.30 & -0.60\\
K00028.01 & 2.050 & 210.88 & 0.007 & 0.000 & 0.000 & 0.000 & 0.850 & 0 & 1 & 0 & 0 & 0.77 & 1.00 & 11200.0 & -1.19 & -0.60\\
K00031.01 & 0.926 & 69.95 & 0.011 & 0.002 & 0.002 & 0.007 & nan & 0 & 2 & 0 & 0 & 1.00 & 1.00 & 135.5 & -1.29 & -0.51\\
K00033.01 & 0.732 & 13.89 & 0.007 & 0.001 & 0.004 & 0.001 & 1.000 & 0 & 2 & 1 & 0 & 1.00 & 0.33 & 13.8 & -1.30 & -0.52\\
K00041.01 & 12.816 & 2.53 & 0.996 & 1.000 & 0.999 & 1.000 & 0.000 & 0 & 1 & 0 & 0 & 0.41 & 0.78 & 68.6 & -1.03 & -0.45\\
K00041.02 & 6.887 & 1.54 & 0.994 & 0.999 & 1.000 & 1.000 & 0.000 & 0 & 1 & 0 & 0 & 0.86 & 0.88 & 27.5 & -1.05 & -0.45\\
K00041.03 & 35.333 & 1.82 & 0.998 & 1.000 & 1.000 & 1.000 & 0.000 & 0 & 1 & 0 & 0 & 0.46 & 0.88 & 18.0 & -1.05 & -0.44\\
K00042.01 & 17.834 & 3.14 & 0.859 & 0.931 & 0.899 & 0.897 & 0.000 & 2 & 0 & 0 & 0 & nan & 0.00 & 130.6 & -1.43 & -0.47\\
\vdots & \vdots &\vdots &\vdots &\vdots &\vdots &\vdots &\vdots &\vdots &\vdots &\vdots &\vdots &\vdots &\vdots &\vdots &\vdots &\vdots \\\hline
\end{tabular}
\begin{tablenotes}
\item [1] Adjusted using Gaia DR2 stellar radius from \citet{Berger:2018il}
\item [2] Planet probability including prior information
\item [3] NASA Exoplanet Archive Astrophysical FPP table values for DR25. Low values indicate increased chance of KOI being a planet, opposite to the GPC and other model values.
\item [4] \citet{Berger:2018il}
\item [5] \citet{Ziegler:2018ei} accounting for source brightness and TCE transit depth.
\item [6] \citet{Bryson:tz}
\end{tablenotes}
\end{threeparttable}
\end{table}
\end{landscape}

\end{document}